\documentclass[prd,12pt,aps,superscriptaddress,preprintnumbers,showpacs,tightenlines,floatfix,nofootinbib,amssymb]{revtex4}
\usepackage{epsfig}

\newcommand{\eff}{{\mathrm{eff}}}

\newcommand{\mon}{{\mathrm{mon}}}
\def\absc{(1-\vert c(s,\mu)\vert^2)^{1/2}}

\newcommand{\cD}{{\cal D}}

\newcommand{\be}{\begin{equation}}
\newcommand{\ee}{\end{equation}}
\newcommand{\beqn}{\begin{eqnarray}}
\newcommand{\eeqn}{\end{eqnarray}}
\newcommand{\eq}[1]{(\ref{#1})}

\newcommand{\Kanazawa}{\affiliation{Institute for Theoretical Physics,
Kanazawa University, Kanazawa 920-1192, Japan}}
\newcommand{\ITEP}{\affiliation{Institute of Theoretical and
Experimental Physics, B.Cheremushkinskaya 25, Moscow, 117259, Russia}}
\newcommand{\RIKEN}{\affiliation{RIKEN, Radiation Laboratory, Wako 351-0158, Japan}}

\begin{document}

\title{Entropy of spatial monopole currents in pure SU(2) QCD at finite temperature}

\author{M.N.~Chernodub}\ITEP
\author{Katsuya~Ishiguro}\Kanazawa\RIKEN
\author{Tsuneo~Suzuki}\Kanazawa\RIKEN

\preprint{KANAZAWA/2005-03}
\preprint{ITEP-LAT/2005-05}

\begin{abstract}
We study properties of space--like monopole trajectories in the Maximal
Abelian gauge of quenched SU(2) QCD at the finite temperature.
We concentrate on infrared monopole clusters which are
responsible for the confinement properties of the theory. We determine
numerically the effective action of the monopoles projected onto the
three-dimensional time--slice. Then we derive the length distributions
of the monopole loops and fix their entropy.
\end{abstract}


\date{March 18, 2005}

\maketitle

\section{Introduction}
\label{one}
The interest in the Abelian monopoles in the non--Abelian gauge theories is
motivated by a central role of these objects in the dual
superconductor mechanism~\cite{DualSuperconductor} of color confinement.
The Abelian monopoles can be considered as particular configurations of
gluon fields with magnetic quantum numbers. In pure non--Abelian gauge theories
the Abelian monopoles do not exist at the classical level. However, these topological
defects can successfully be identified given a dynamical configuration of the
gluon fields in a particular Abelian gauge~\cite{AbelianProjections}. There are
many Abelian gauges among which the most popular one is the Maximal Abelian (MA)
gauge~\cite{kronfeld}. In this gauge the off-diagonal gluon fields are
suppressed and short--ranged contrary to the diagonal (Abelian)
gluon fields~\cite{ref:propagators}. There are many numerical experiments
confirming that Abelian degrees of freedom are responsible for the
confinement of color (for a review, see Ref.~\cite{Reviews}).
In particular, it was observed in Refs.~\cite{AbelianDominance,shiba:string}
that the tension of the chromoelectric string is dominated by the
Abelian monopole contributions. Moreover, the monopole condensate
-- which guarantees the formation of the chromoelectric string between the
quarks -- exists  in the confinement phase and disappears in the deconfinement
phase~\cite{shiba:condensation,MonopoleCondensation}.

The trajectories of the Abelian monopoles form two different types of clusters.
A typical configuration contains a lot of finite-sized clusters and one
large percolating cluster~\cite{ivanenko,ref:kitahara}.
The percolating cluster (or, the infrared (IR) cluster) occupies the whole lattice
while the sizes of the other clusters have an ultraviolet nature.
The monopole condensate corresponds to the so-called percolating (infrared)
cluster of the monopole trajectories.  The tension of the
confining string gets a dominant contribution from the IR monopole
cluster~\cite{ref:kitahara} while the finite-sized ultraviolet (UV) clusters
do not play any role in the confinement. Various properties of the UV
and IR monopole clusters were investigated previously in
Refs.~\cite{ref:kitahara,ref:clusters,IshiguroSuzuki}.

At high temperatures the Abelian monopoles become static. In the high temperature phase
the IR monopole cluster disappears~\cite{ivanenko,ref:kitahara} and, consequently, the
confinement of the static quarks is lost. Since the static currents do not play
any role in confinement we concentrate below on the spatial components of the IR
monopole cluster. We investigate the action, the length distribution and the entropy
of spatial components of the infrared monopole clusters. We follow
Ref.~\cite{IshiguroSuzuki} where energy and entropy of the monopole currents were studied at
zero temperature. Our preliminary results were reported in
Ref.~\cite{IshiguroSuzuki:spatial:Lattice}.

The plan of the paper is as follows. In Section~\ref{sec:model} we describe the model and
provide the description of the monopole currents. The details of numerical simulations are
also given in this Section. Section~\ref{sec:action} is devoted to the investigation of the
Abelian monopole action obtained by the inverse Monte-Carlo method for the clusters of the
spatially projected Abelian monopoles. In Section~\ref{sec:length} we study the length
distributions of the infrared clusters of the spatially projected monopole clusters.
The knowledge of the monopole action and cluster distribution allows us to calculate the
entropy of the spatial monopole currents which is discussed in Section~\ref{sec:entropy}.
Our conclusions are presented in the last Section.

\section{Model}
\label{sec:model}

We study pure SU(2) QCD with the standard Wilson lattice action for gluon fields,
\beqn
S(U) = - \frac{\beta}{2} {\mathrm{Tr}} \sum_P U_P\,,
\eeqn
where $\beta$ is the coupling constant, the sum goes over all plaquettes
of the lattice, and
$U_P \equiv U_{s,\mu\nu} = U_{s,\mu}U_{s+\hat\mu,\nu}
U^\dagger_{s+\hat\nu,\mu} U^\dagger_{s,\nu}$ is the SU(2) plaquette constructed
from link fields, $U_{s,\mu}$. We work in the MA
gauge~\cite{kronfeld} defined by the maximization of the lattice functional
\beqn
R = \sum_{s,\hat\mu}{\mathrm{Tr}}\Big(\sigma_3 \widetilde{U}(s,\mu)
\sigma_3 \widetilde{U}^{\dagger}(s,\mu)\Big)\,,
\label{R}
\eeqn
with respect to the gauge transformations
$U(s,\mu) \to \widetilde{U}(s,\mu)=\Omega(s)U(s,\mu)\Omega^\dagger(s+\hat\mu)$.
In the continuum limit the local condition of maximization~\eq{R} can be
written in terms of the differential
equation, $(\partial_{\mu}+igA_{\mu}^3)(A_{\mu}^1-iA_{\mu}^2)=0$.
Both this condition and the functional \eq{R} are invariant under
residual U(1) gauge transformations, $\Omega^{\mathrm{Abel}}(\omega)
= {\mathrm{diag} (e^{i \omega(s)},e^{- i \omega(s)})}$.

After the Abelian gauge is fixed we perform the projection of the non-Abelian
gauge fields, $U_{s,\mu}$, onto the Abelian ones, $u_{s,\mu}$:
\beqn
 \widetilde{U}(s,\mu) = \left( \begin{array}{cc}
         \absc        & -c^*(s,\mu) \\
                  c(s,\mu) &  \absc
\end{array} \right)
\left( \begin{array}{cc}
u(s,\mu) & 0 \\
0 & u^*(s,\mu)
\end{array} \right),
\label{eq:field:decomposition}
\eeqn
where $c(s,\mu)$ corresponds to the charged (off-diagonal) matter fields.

As we have discussed above, the dominant information about the confinement
properties of the theory is located in the monopole configurations which
are identified with the help of the Abelian phases of the
diagonal fields, $\theta_{s,\mu}$.
The Abelian field strength $\theta_{\mu\nu}(s)\in(-4\pi,4\pi)$ is defined
on the lattice plaquettes by a link angle $\theta(s,\mu)\in[-\pi,\pi)$
as $\theta_{\mu\nu}(s)=\theta(s,\mu)+
\theta(s+\hat\mu,\nu)-\theta(s+\hat\nu,\mu)-\theta(s,\nu)$.
The field strength $\theta_{\mu\nu}(s)$ can be decomposed into two parts,
\beqn
\theta_{\mu\nu}(s)= \bar{\theta}_{\mu\nu}(s) +2\pi m_{\mu\nu}(s)\,,
\label{eq:field:separation}
\eeqn
where $\bar{\theta}_{\mu\nu}(s)\in [-\pi,\pi)$ is interpreted as
the electromagnetic flux through the plaquette
and $m_{\mu\nu}(s)$ can be regarded as a number of the Dirac
strings piercing the plaquette.

The elementary monopole current can conventionally be constructed using
the DeGrand-Toussaint\cite{degrand} definition:
\beqn
k_{\mu}(s) & = & \frac{1}{2}\epsilon_{\mu\nu\rho\sigma}
\partial_{\nu}m_{\rho\sigma}(s+\hat{\mu}),
\label{eq:monopole:definition}
\eeqn
where $\partial$ is the forward lattice derivative. The monopole current is defined
on a link of the dual lattice and takes the values $0, \pm 1, \pm 2$. Moreover the
monopole current satisfies the conservation law automatically,
\beqn
\partial'_{\mu}k_{\mu}(s)=0\,,
\eeqn
where $\partial'$ is the backward derivative on the dual lattice.

The monopole current~\eq{eq:monopole:definition} corresponds to the monopole
charge defined on the scale of the elementary lattice spacing, $a$. Obviously,
the scale $a$ becomes smaller as we approach the continuum limit. In order to
study the properties of the monopoles at fixed {\it physical} scales
we use the so--called extended monopoles~\cite{ivanenko}. The $n^3$ extended
monopole is defined on a coarse sublattice with the lattice spacing $b=na$.
Thus the construction of the extended monopoles corresponds to a block--spin
transformation of the monopole currents with the scale factor $n$,
\beqn
k_{\mu}^{(n)}(s) = \sum_{i,j,l=0}^{n-1}k_{\mu}(n s+(n-1)\hat{\mu}+i\hat{\nu}
     +j\hat{\rho}+l\hat{\sigma})\,.
\label{eq:blocking}
\eeqn

Since the time--like monopole currents are not essential for the confinement
properties we concentrate on the spatial components of the currents. Namely,
we investigate spatially projected currents,
\beqn
K_i^{(n)}(\vec s) = \sum^{L_t-1}_{s_4=0}\, k_i^{(n)}(s,s_4)\,,\,\, i=1,2,3\,,
\eeqn
which are integer--valued and closed.

Technically, we generate 2000-10000 configurations of  the SU(2) gauge field,
$U$, for $\beta=2.3 \sim 2.6$ on the lattices $L_s^3\times L_t$,
with $L_s = 24,32,48,72$ and $L_t=4,6,8,12,16$. The number of
generated configuration depends on the value of $\beta$ and
lattice volume. We fix the gauge with the help of the usual iterative algorithm.
In this paper we used the same methods as in the zero--temperature case
studied in Ref.~\cite{IshiguroSuzuki}. Thus we refer an interested reader
to Ref.~\cite{IshiguroSuzuki} for a more detailed description of the numerical
procedures. Below we concentrate on the description of the numerical results.

\section{Monopole action}
\label{sec:action}

In what follows we discuss an effective model of the monopole currents corresponding
to pure SU(2) QCD. Formally, we get this effective model through the gauge
fixing procedure applied to the original model. Then we integrate out
all degrees of freedom but the monopole ones. An effective monopole action
is related to the original non-Abelian action $S[U]$ as follows:
\beqn
Z  &=& \int \cD U \, \delta(X) \Delta_{FP}(U)\, e^{- S[U]}
= \Bigl( \prod_{s, \mu} \sum_{k_\mu(s) = -\infty}^{\infty} \Bigr)
        \Bigl( \prod_s \delta_{ \partial_{\mu}^{\prime} k_\mu (s), 0} \Bigr)
        e^{-S^{\mon}_{\eff}[k]}\,.
\label{eq:Z1}
\eeqn
We omit irrelevant constant terms in front of the partition function.
The term $\delta(X)$ represents the gauge-fixing condition
and $\Delta_{FP}(U)$ is the corresponding Faddeev-Popov determinant.
As we have discussed above, the MA gauge fixing condition is given by
a maximization of the functional~\eq{R} and therefore the
local condition $X=0$, implied in Eq.~\eq{eq:Z1}, is used here as a formal
simplified notation.

Numerically, the monopole action of the $3D$ projected IR monopole clusters
can be defined using the inverse Monte--Carlo method~\cite{shiba:condensation}.
The action is represented in a truncated form~\cite{shiba:condensation,chernodub}
as a sum of the $m$--point ($m \ge 2$) operators~$S_i$:
\beqn
S_{\mon}[K] = \sum\nolimits_i f_i S_i [K]\,,
\label{eq:monopole:action}
\eeqn
where $f_i$ are the coupling constants. Following Ref.~\cite{IshiguroSuzuki}
we adopt only the two--point interactions in the monopole action, $S_i \sim K_{i}(s) K_{j}(s')$.

\begin{figure}[!thb]
\begin{center}
\begin{tabular}{cc}
\includegraphics[scale=0.45,clip=false]{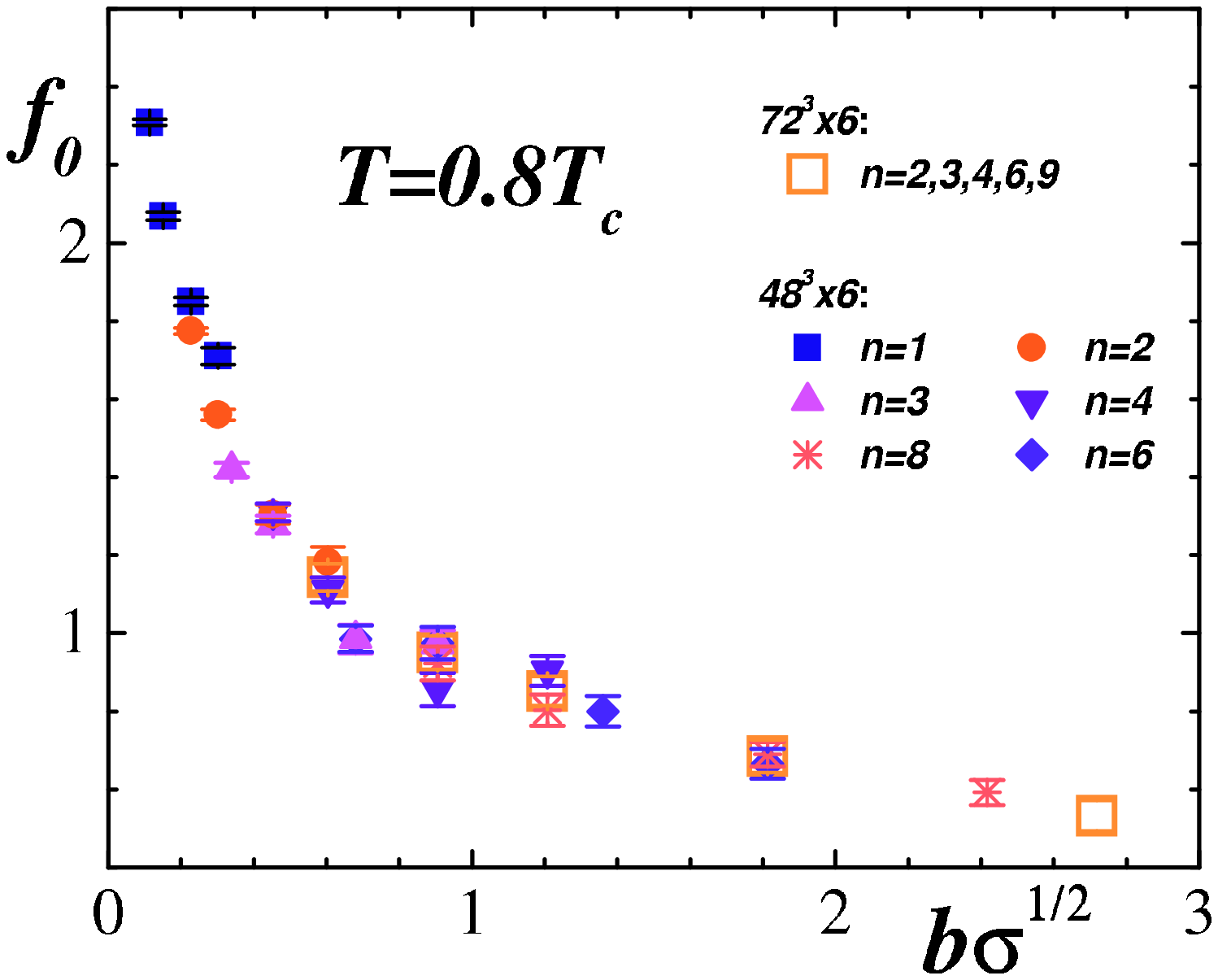} \hspace{5mm} &
\includegraphics[scale=0.45,clip=false]{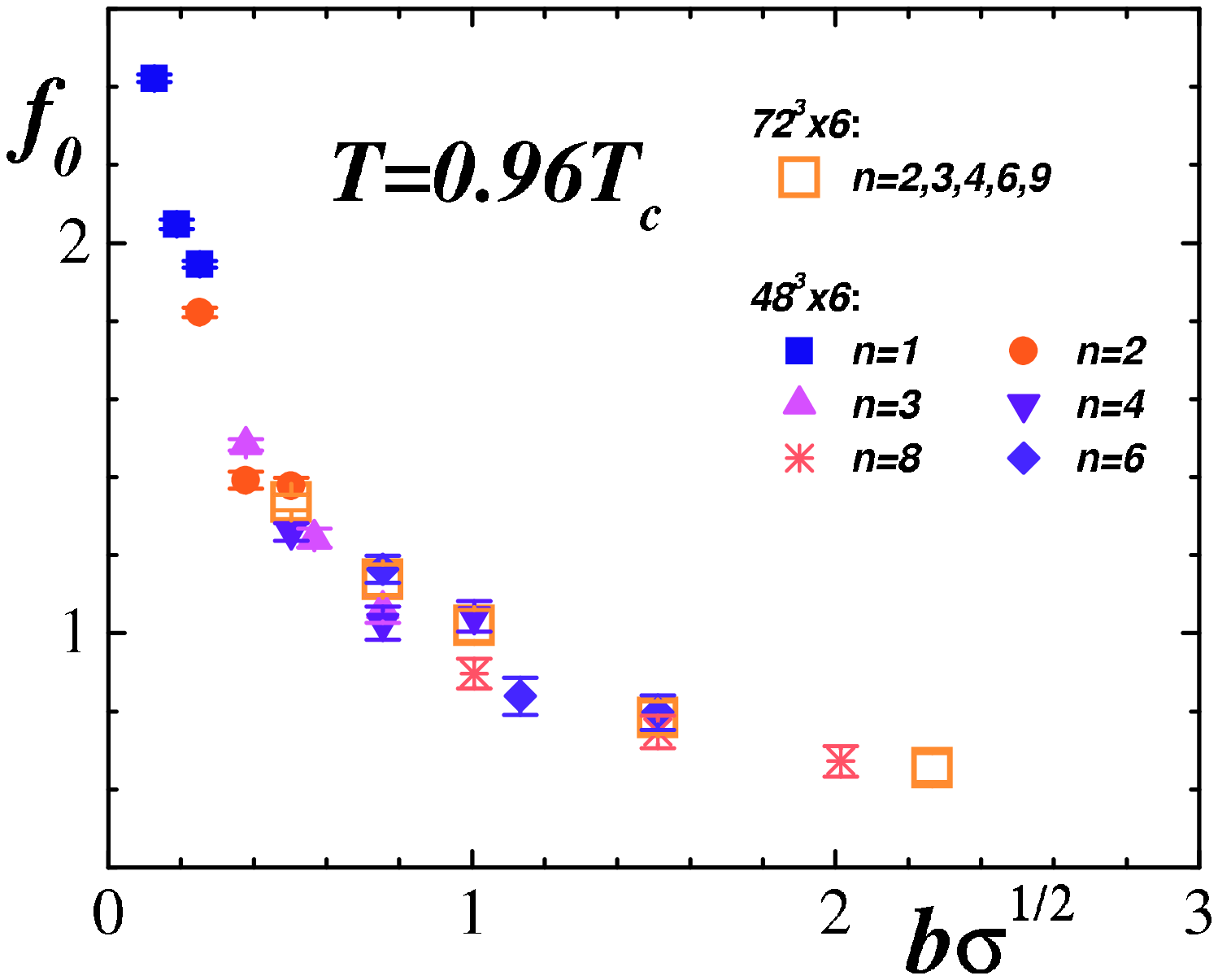} \\
(a) & (b)
\end{tabular}
\end{center}
\vspace{-4mm}
\caption{The coefficient $f_0$ of the monopole action~\eq{eq:length:proportionality}
{\it vs.} the scale parameter $b$ for the lattice sizes $L_s^3\times 6$, $L_s=48,72$
and blocking factors, $n=1 \dots 9$, at temperatures (a) $T=0.8\,T_c$
and (b) $T=0.96\,T_c$.}
\label{fig:f0}
\end{figure}

Similarly to the $4D$ case we find that the monopole action of the spatially projected
currents is proportional with a good accuracy to the length $L[K]$ of the monopole
loop $K$,
\beqn
S_{\mon}[K] \backsimeq f_0 L[K] + const\,.
\label{eq:length:proportionality}
\eeqn
The important property of the monopole action is that the couplings $f_i$ are the
functions of the scale $b=na$, Eq.~\eq{eq:blocking}, at which the monopole
charge is defined. To illustrate this fact we show the dependence of the
coupling constant $f_0$ on $b = n\, a(\beta)$ in Figure~\ref{fig:f0}.

{}From Figure~\ref{fig:f0} one observes the almost perfect scaling: the parameter $b$ does not depend on the
parameters $n$ and $a$ separately. The action is near to the renormalized trajectory
which corresponds to the continuum effective action. Moreover, the result does not depend on the
spatial extension of the lattice, $L_s$. Thus the action of the spatially projected
monopole current shows the scaling similarly to the action of the unprojected
monopoles~\cite{shiba:condensation,chernodub}.

\section{Length distribution}
\label{sec:length}

\begin{figure}[!htb]
\begin{center}
\vspace{3mm}
\begin{tabular}{cc}
\includegraphics[scale=0.45,clip=true]{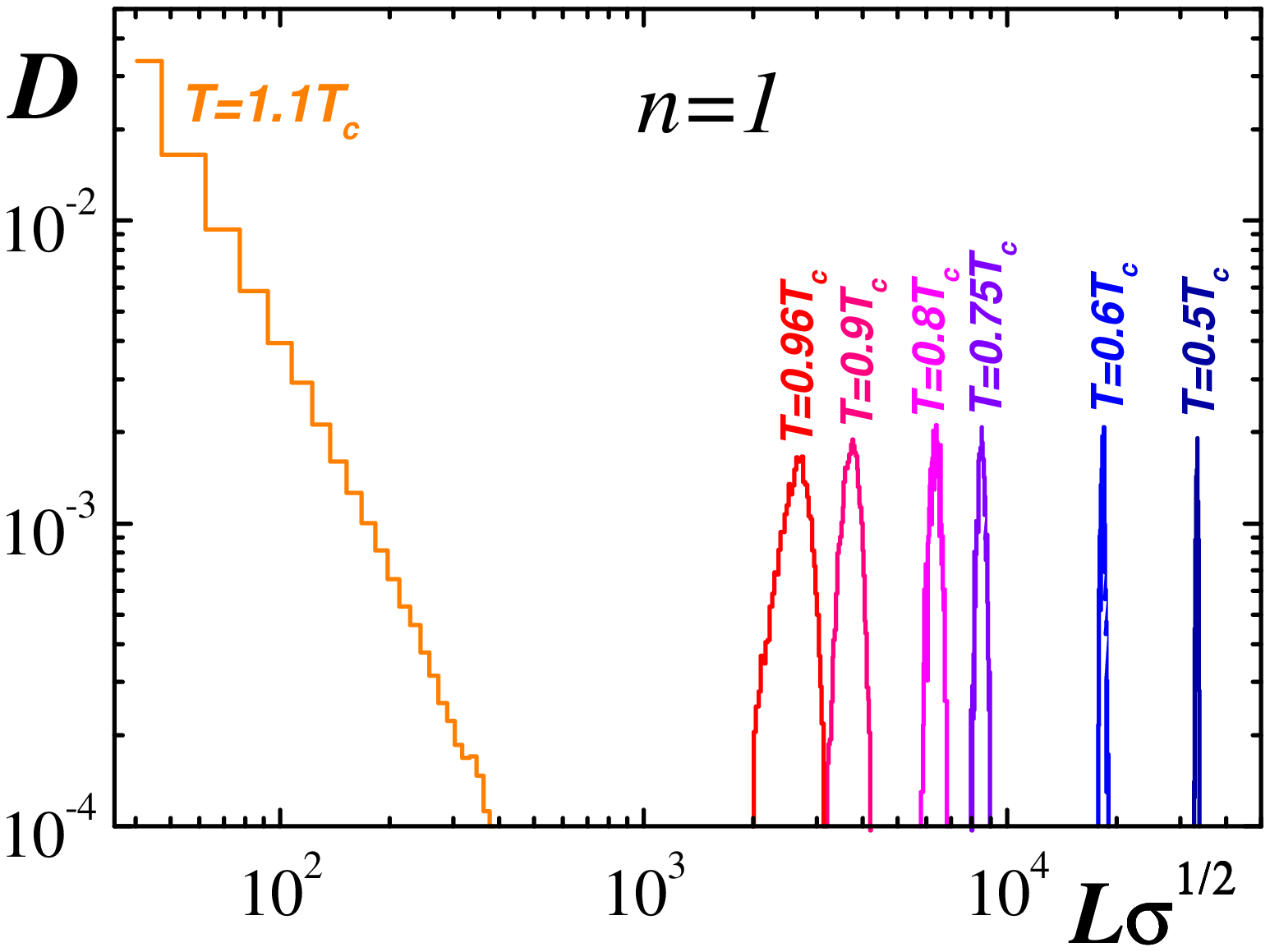} \hspace{5mm} &
\includegraphics[scale=0.45,clip=true]{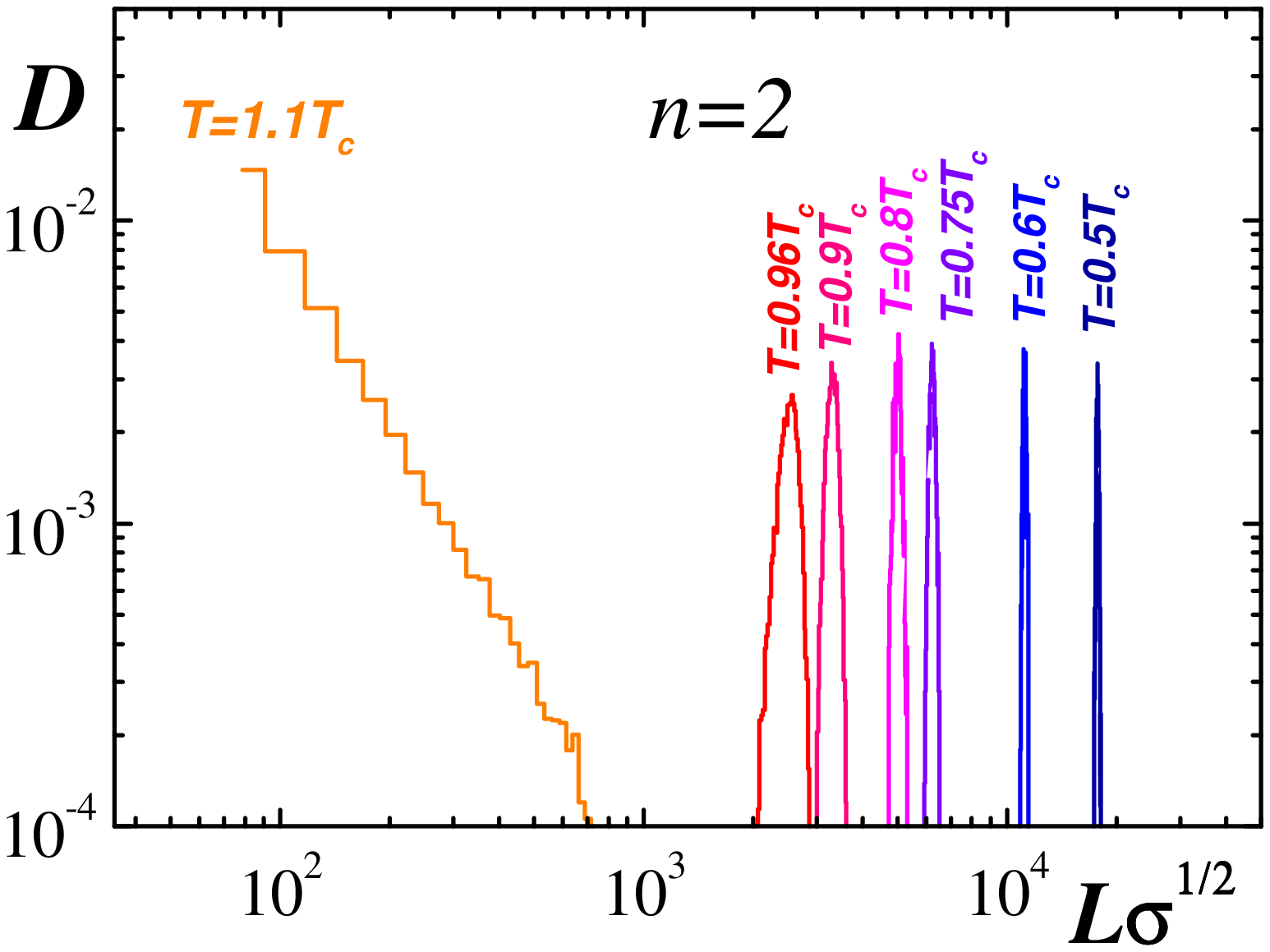} \\
(a) & (b)
\end{tabular}
\end{center}
\vspace{-3mm}
\caption{The distributions of the spatial monopole currents at various temperatures.
For the low temperatures, $T < T_c$, the UV--part of the distributions is not shown.}
\label{fig:Distr}
\end{figure}

The length distribution of the spatially--projected monopole clusters
is shown in Figure~\ref{fig:Distr}.
In the confinement phase, $T<T_c$,
only the infrared part of the distribution is shown. One can see that
the length (in physical units) of the monopole trajectory belonging to
the percolating cluster becomes shorter as the temperature increases.
This fact is expected because the monopole condensate is "evaporating"
as temperature increases towards the transition point, and, therefore,
the infrared part of the monopole currents should be more and more diluted.

At $T>T_c$ the percolating cluster of the spatially projected currents
disappears, and, consequently, the confinement of quarks is lost.
The behavior of the elementary and blocked currents is qualitatively the same.
According to Ref.~\cite{IshiguroSuzuki} the length of the $4D$ IR monopole currents
in the finite volume $V$ is distributed with the Gaussian law, which is
in the finite-temperature case can be formulated as follows:
\beqn
D^{IR}(L) \propto \exp\{ - \alpha(b,V) L^2 + \gamma(b,T) L\}\,.
\label{eq:IR:distr:two}
\eeqn
The length distribution function, $D(L)$, is proportional to the
weight with which the particular trajectory of the length $L$ contributes to
the partition function. In Eq.~\eq{eq:IR:distr:two} we neglect a
power-law prefactor, $1/L^\tau$ with $\tau \sim 3$, which is essential for
the distribution of the infrared clusters.

The Gaussian form of the distribution \eq{eq:IR:distr:two} means that the
clusters have the typical length
\beqn
L_{max} = \gamma(b,T) \slash 2 \, \alpha(b,V)\,,
\label{eq:Lmax}
\eeqn
where $V$ is the three--dimensional volume.
The coefficient $\alpha$ plays a role of the infrared cut--off which emerges
due to the finite volume. In other words, the length of the monopole trajectory in an
infrared cluster is restricted by the lattice boundary. However, since the cluster
is infrared the length of the monopole trajectory in this cluster must be proportional
to the total volume, $L_{max} \propto V$. The linear part of the distribution~\eq{eq:IR:distr:two}
gets contribution from the monopole action and the monopole entropy (we discuss this issue below).
Therefore the coefficient $\gamma$ should not depend on the volume in the thermodynamic limit.
Thus, we expect
\beqn
\alpha(b,V) = A(b) \slash V\,,
\label{eq:A}
\eeqn
where $A(b)$ is a certain function of the scale parameter $b$. One may suggest that the parameter $A$
should not significantly depend on the temperature $T$ since the factor is more kinematical than dynamical.
The temperature influences the dynamical characteristics of the monopoles such as the effective
three-dimensional action. The effective monopole action contributes to the coefficient $\gamma$
and, as a consequence, the temperature influences the projected monopole density via the $\gamma$--coefficient.

Using Eqs.~(\ref{eq:Lmax},\ref{eq:A}) one can obtain that
the monopole density in the infrared cluster is finite in the thermodynamic limit and is given by
the formula
\beqn
\rho_{IR} = \frac{L_{max}}{V} \equiv \frac{\gamma(b,T)}{2 \, A(b)}\,.
\label{eq:density}
\eeqn

\begin{figure}[!htb]
\begin{center}
\vspace{3mm}
\begin{tabular}{cc}
\includegraphics[scale=0.45,clip=true]{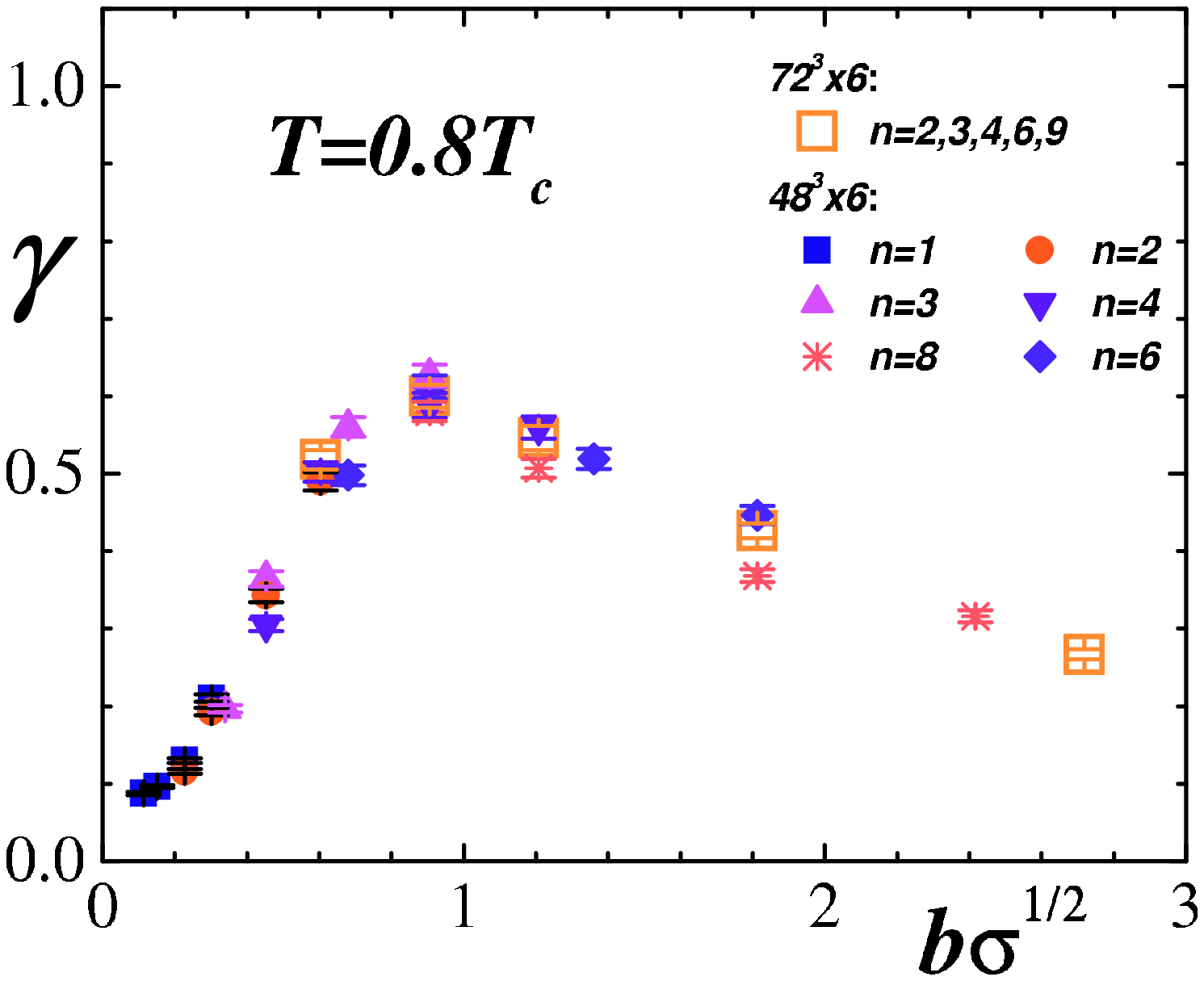} \hspace{5mm} &
\includegraphics[scale=0.45,clip=true]{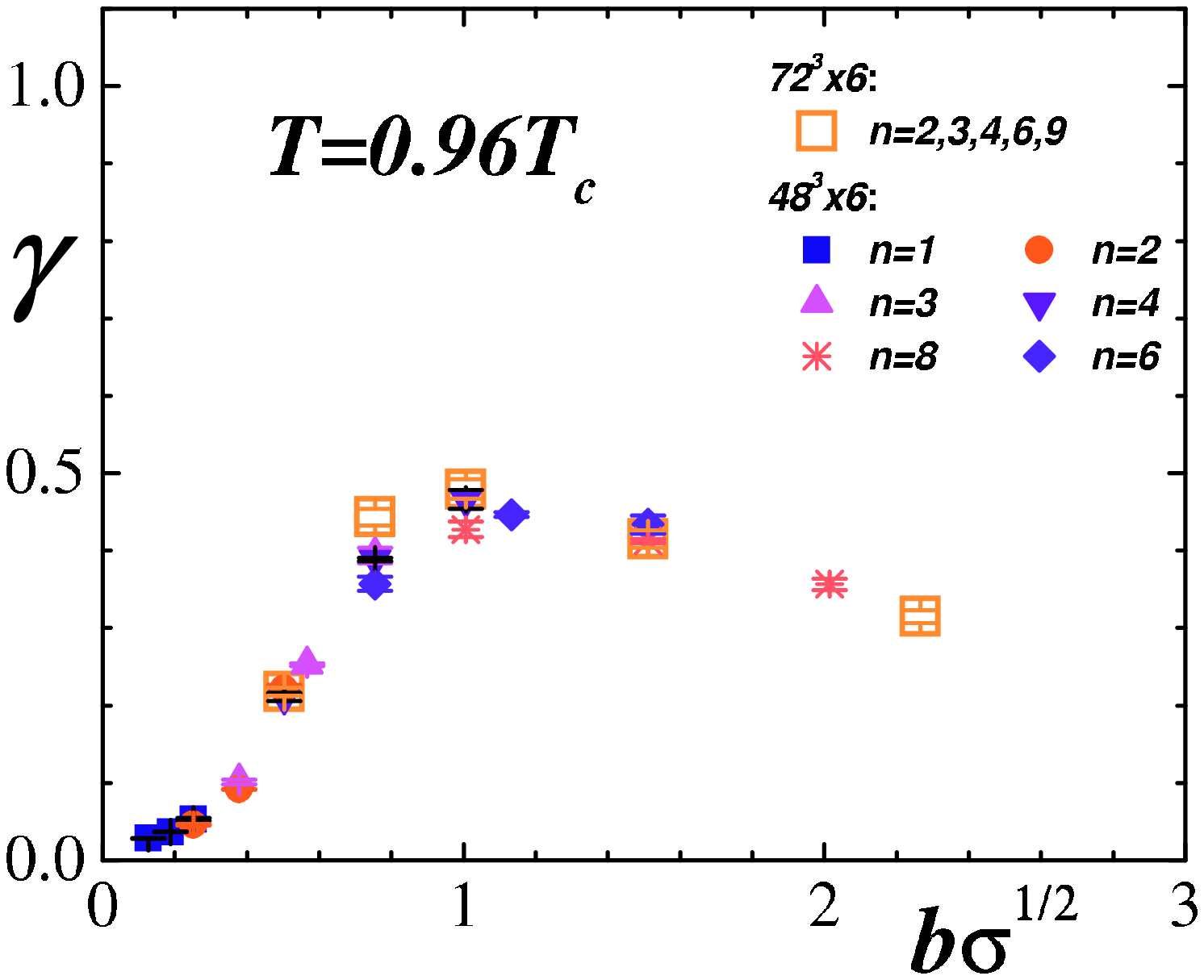} \\
(a) & (b)
\end{tabular}
\end{center}
\vspace{-3mm}
\caption{The same as in Figure~\ref{fig:f0} but for the coefficient $\gamma$
of the distribution~\eq{eq:IR:distr:two}.}
\label{fig:gamma}
\end{figure}

We fit the numerically obtained distributions of
the $3D$ projected currents by the function~\eq{eq:IR:distr:two}
and then use a bootstrap method\footnote{A detailed
description of the corresponding bootstrap method is given in
Ref.~\cite{IshiguroSuzuki}.} to estimate the statistical errors
of the fitting parameters. In Figure~\ref{fig:gamma}
we show the coupling constant $\gamma(b,T)$ as a function of the scale
parameter, $b$, at temperatures $T=0.8 \, T_c$ and $T = 0.96 \, T_c$.
Again, as in the case of the parameter $f_0$, Figure~\ref{fig:f0}, we
observe the volume independence and the $b$--scaling of the results.
In a small $b$--region we find that $\gamma \propto b^\eta$
with $\eta \sim 3$ for low temperatures, $T \sim 0.5 T_c$,
whereas $\eta \sim 2$ for $T \to T_c$. The data show a good $b$--scaling
and also is independent of the volume similarly to the monopole action.

\begin{figure}[!htb]
\begin{center}
\vspace{3mm}
\begin{tabular}{cc}
\includegraphics[scale=0.45,clip=true]{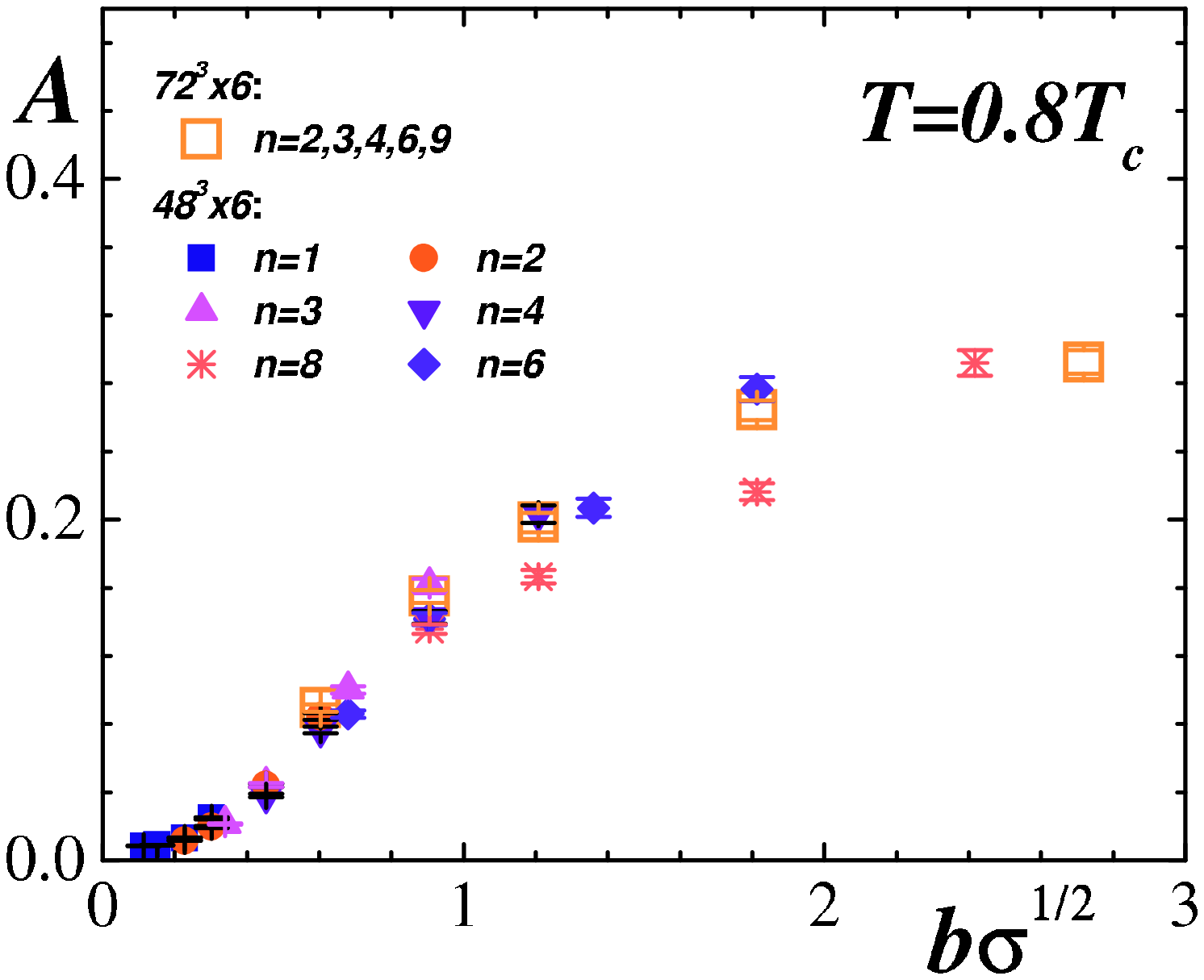} \hspace{5mm} &
\includegraphics[scale=0.45,clip=true]{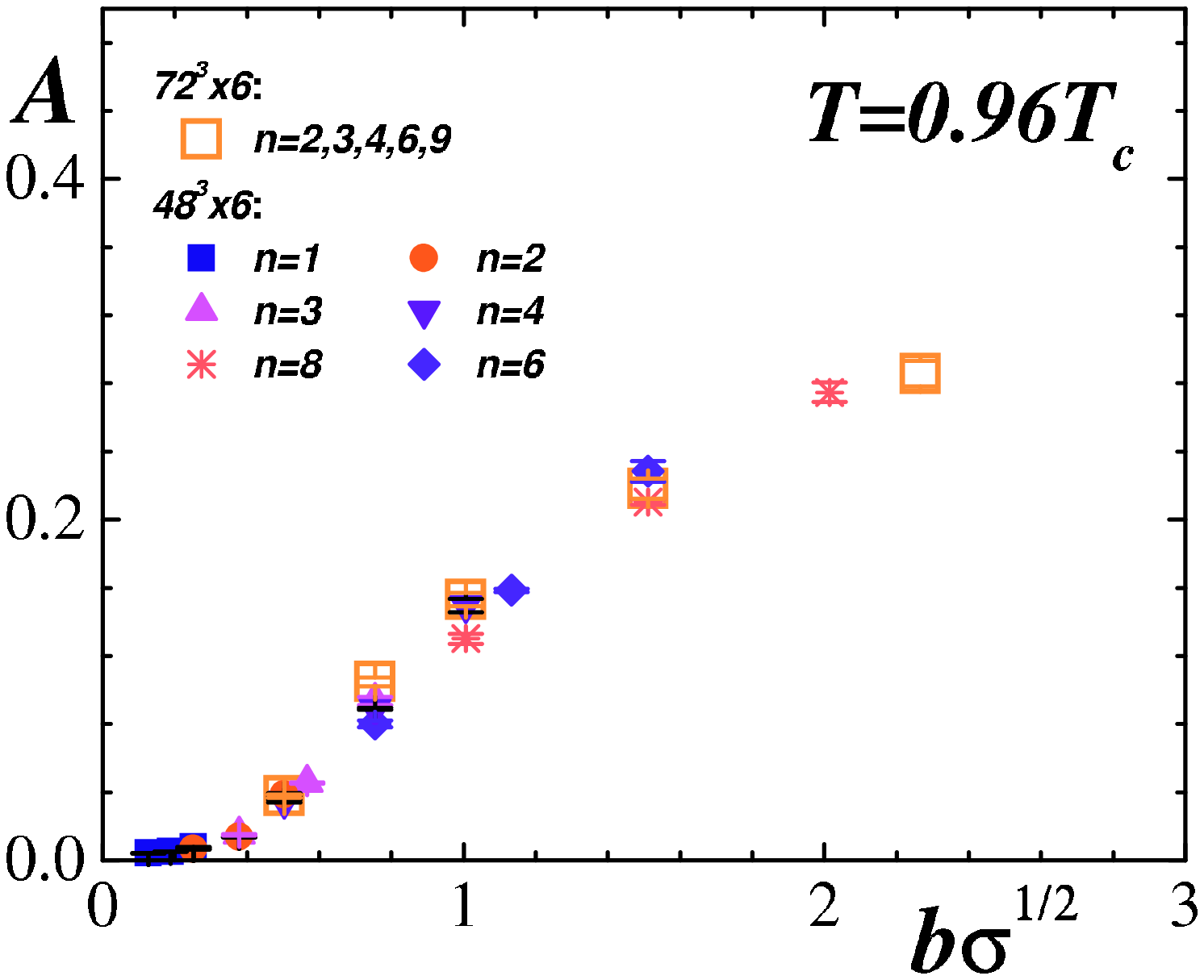} \\
(a) & (b)
\end{tabular}
\end{center}
\vspace{-3mm}
\caption{The same as in Figure~\ref{fig:f0} but for the ratio~$A(b)$, Eq.~\eq{eq:A}.}
\label{fig:A}
\end{figure}

The numerical values of the parameter $A$ are shown in
Figure~\ref{fig:A}. The parameter $A$ is independent of the lattice volume,
indicating that in the thermodynamic limit the coefficient $\alpha$
in the Gaussian distribution~\eq{eq:IR:distr:two} vanishes.

\begin{figure}[!htb]
\begin{center}
\vspace{3mm}
\begin{tabular}{cc}
\includegraphics[scale=0.45,clip=true]{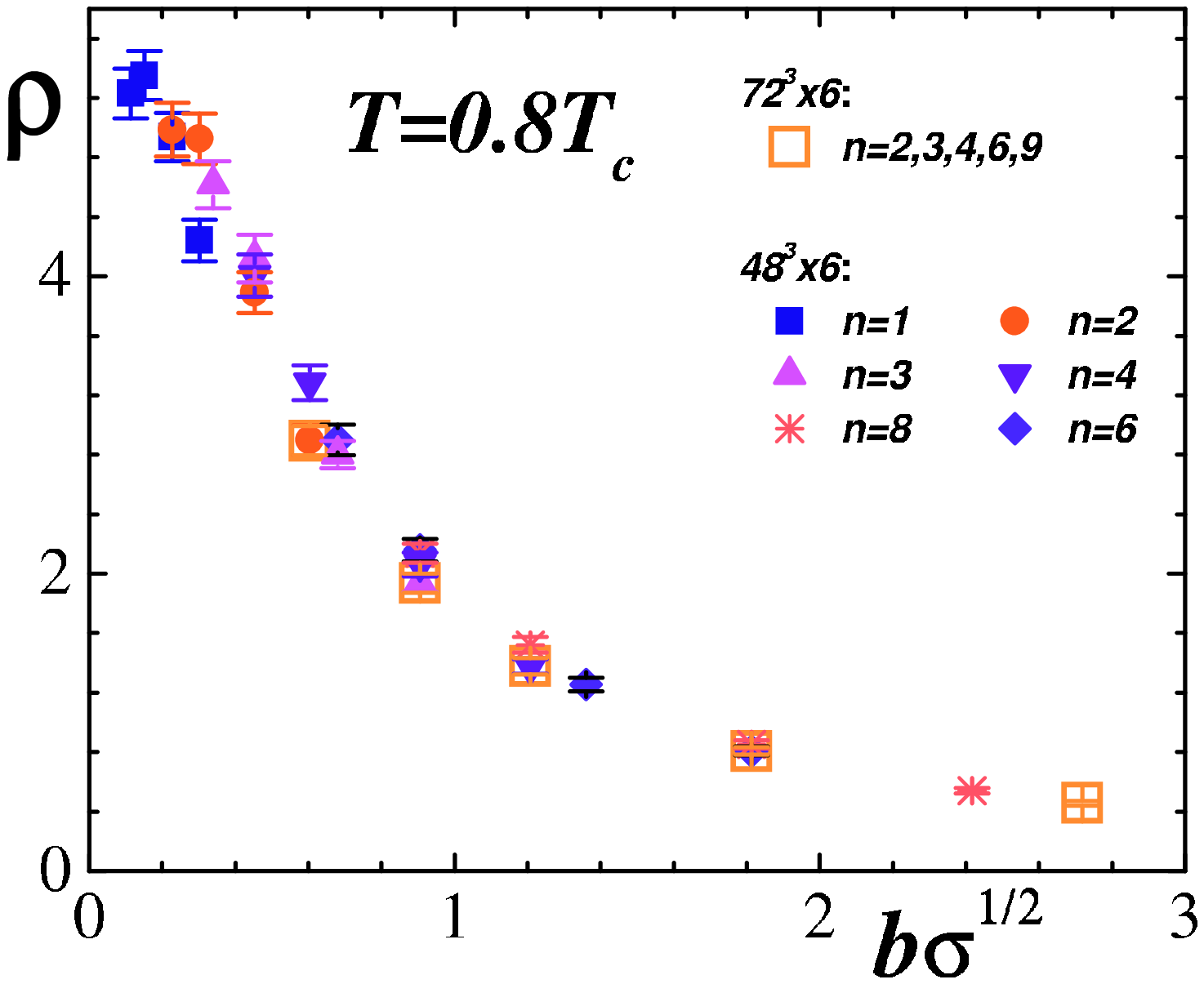} \hspace{5mm} &
\includegraphics[scale=0.45,clip=true]{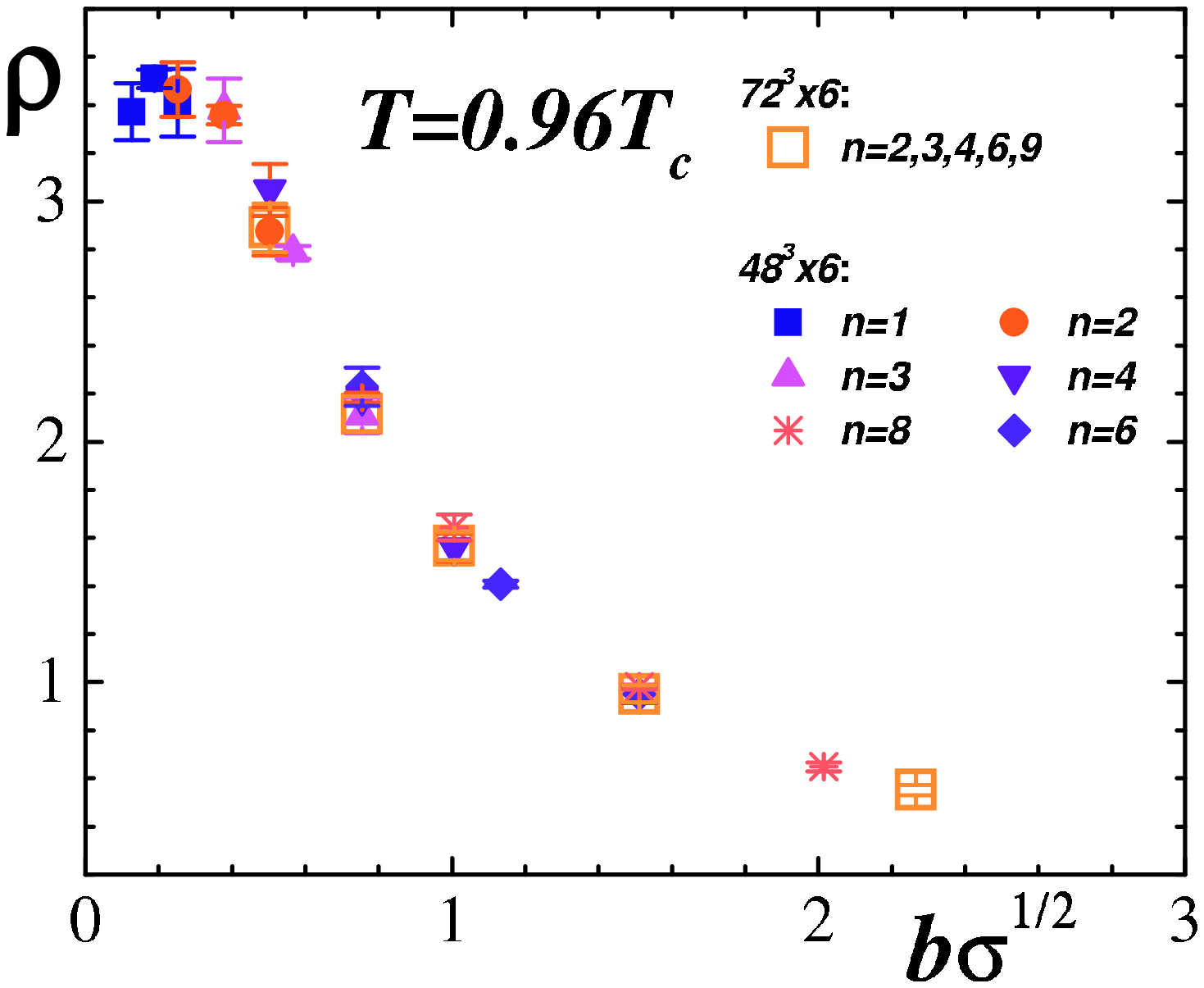} \\
(a) & (b)
\end{tabular}
\end{center}
\vspace{-3mm}
\caption{The same as in Figure~\ref{fig:f0} but for the monopole density~$\rho$ corresponding to the infrared
cluster, Eq.~\eq{eq:density}.}
\label{fig:rho}
\end{figure}

Using Eq.~\eq{eq:density} we calculate the monopole density corresponding to the infrared cluster.
The density is shown in Figure~\ref{fig:rho}.
The density diminishes as the scale factor $b$ increases, while at small $b$ the density
shows a plateau. As the temperature increases the density (at a fixed value of $b$) becomes smaller.

Note that the confining
non--Abelian objects have a finite size (in physical units). These objects are identified as the Abelian monopoles
in the Abelian gauge. The monopoles are detected using the Gauss theorem applied to the magnetic
field coming outside the cube of the size $b^3$. The confining non--Abelian objects have a typical size
which is associated with the size of the monopole core~\cite{ref:rmon}, $r_{mon}\approx 0.05$~fm. If
$b < r_{mon}$ then the monopole cube is too small to detect the charge of much larger monopole and the
monopole density -- measured using the Gauss law -- is vanishingly small. Indeed, one can see that the monopole
density in Figure~\ref{fig:rho} has a tendency to diminish at smaller $b \sqrt{\sigma} \lesssim 0.1$.

\begin{figure}[!thb]
\begin{center}
\vspace{3mm}
\begin{tabular}{cc}
\includegraphics[scale=0.45,clip=true]{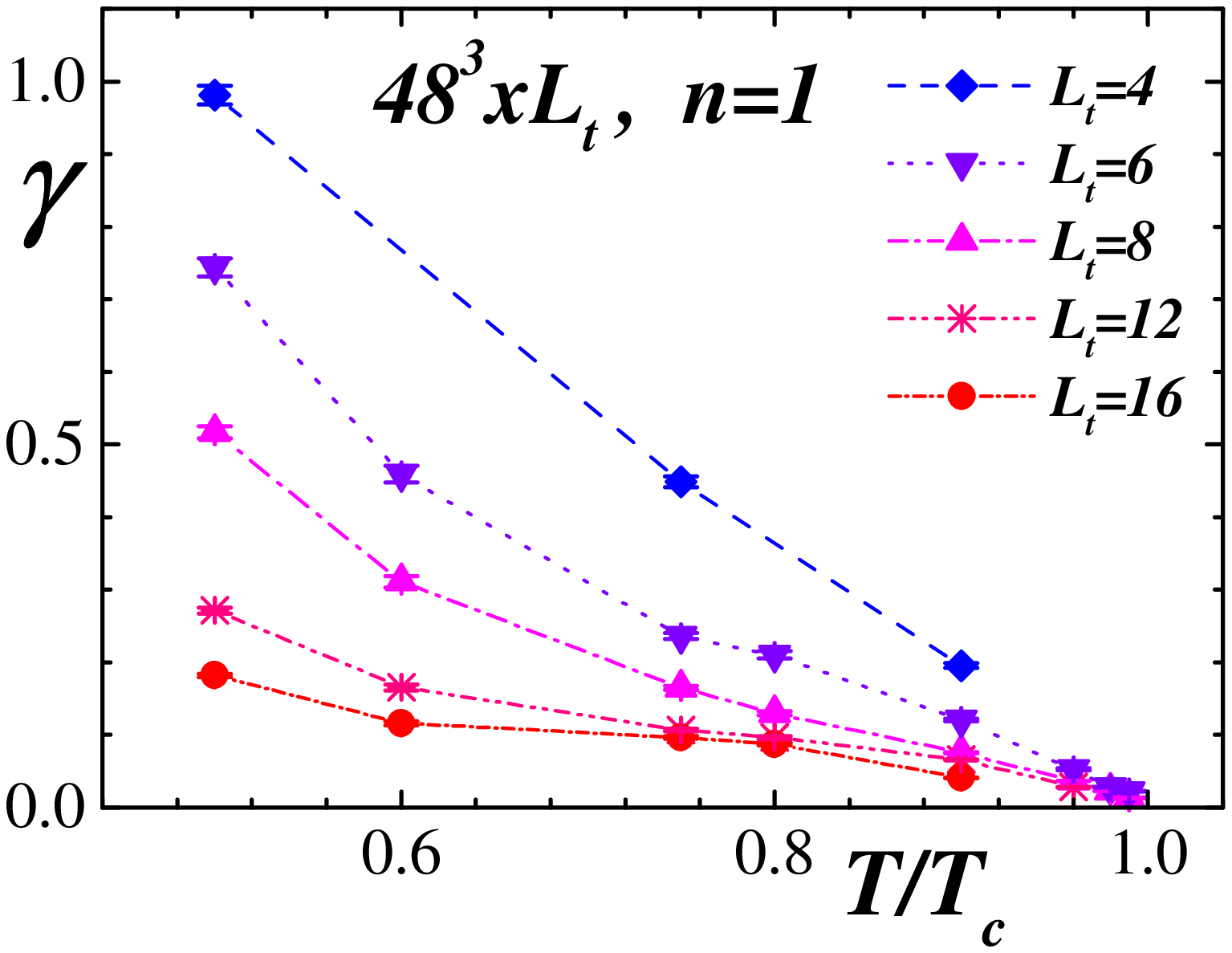} \hspace{5mm} &
\includegraphics[scale=0.45,clip=true]{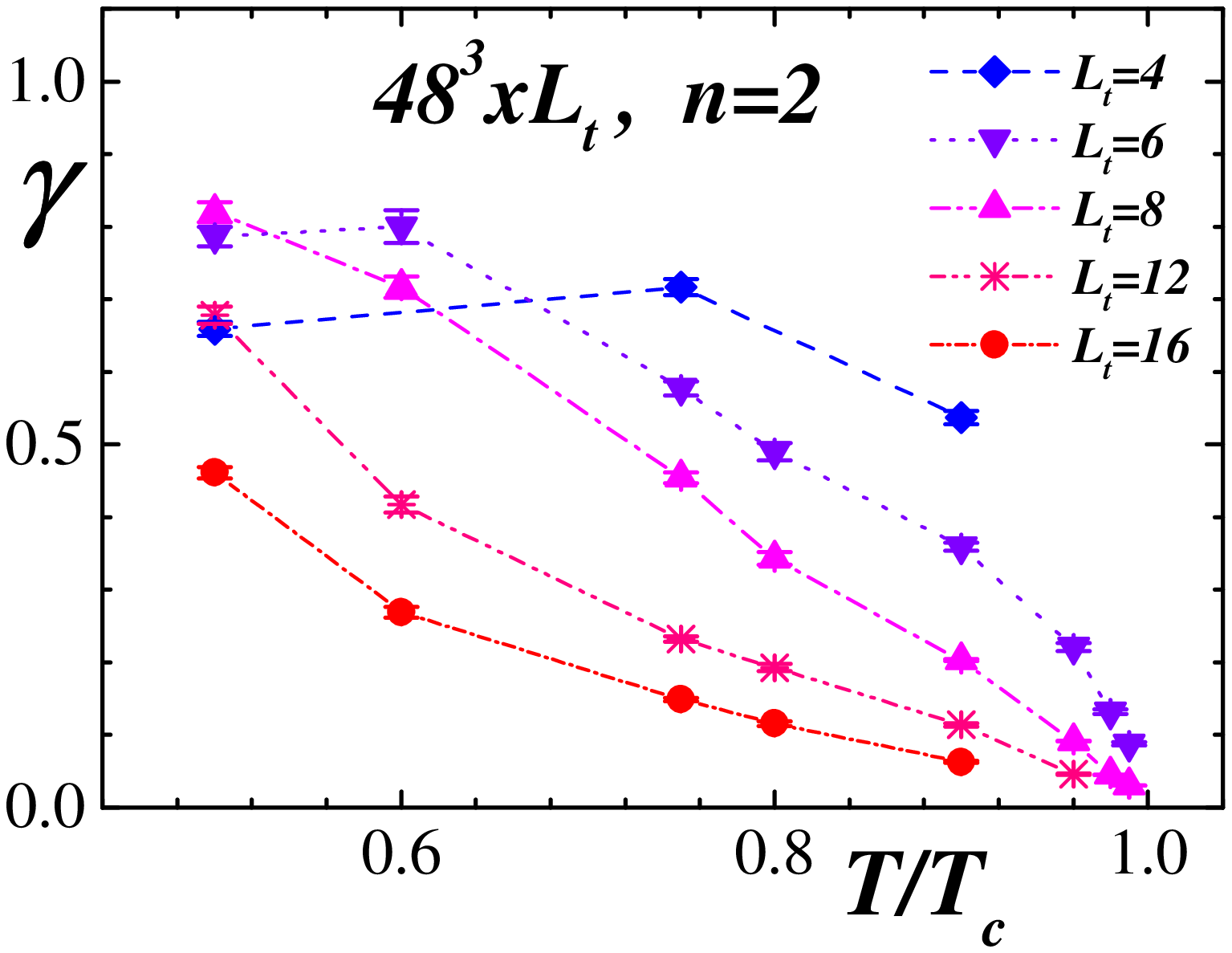} \vspace{-2mm}\\
(a) & (b) \vspace{5mm} \\
\includegraphics[scale=0.45,clip=true]{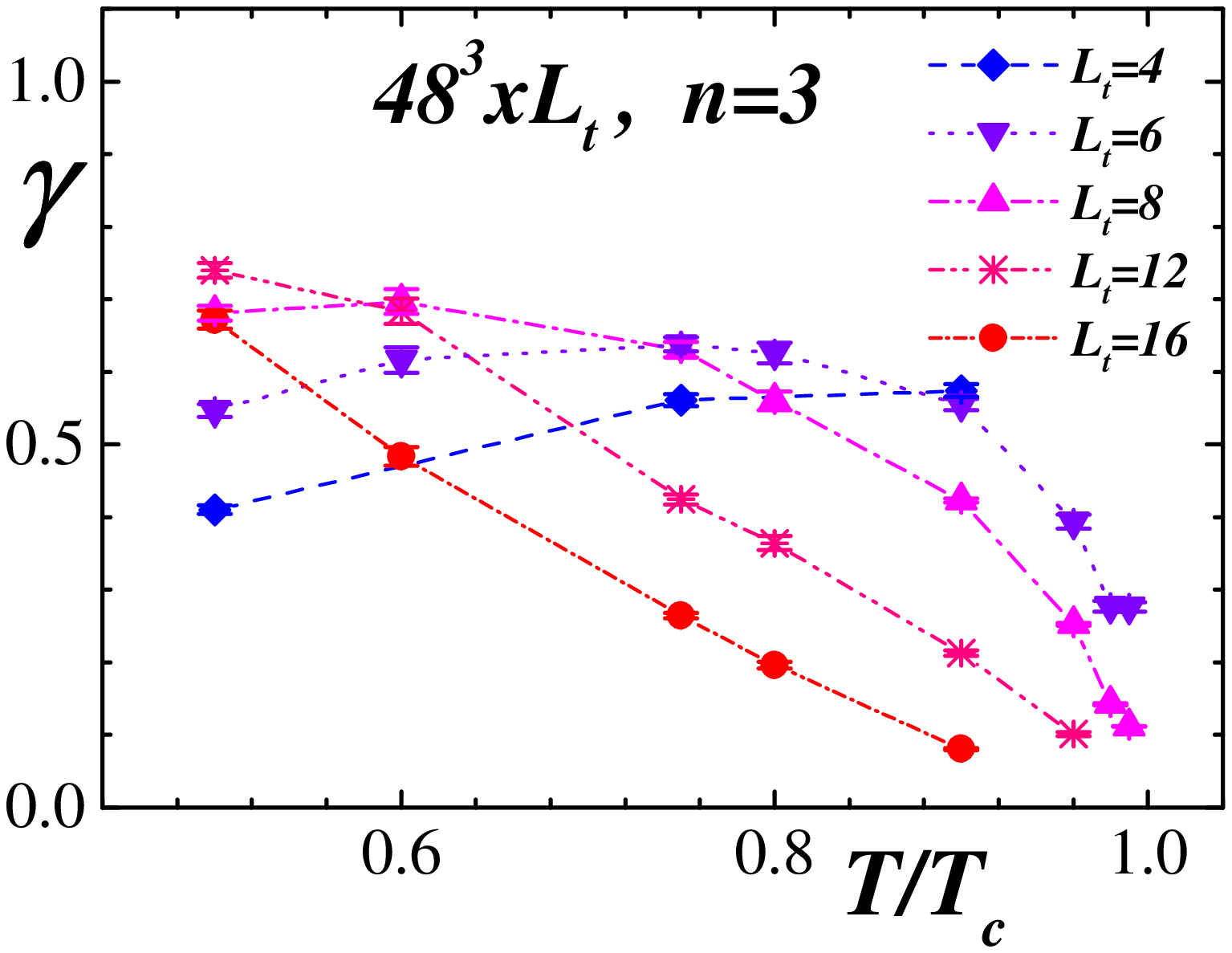} \hspace{5mm} &
\includegraphics[scale=0.45,clip=true]{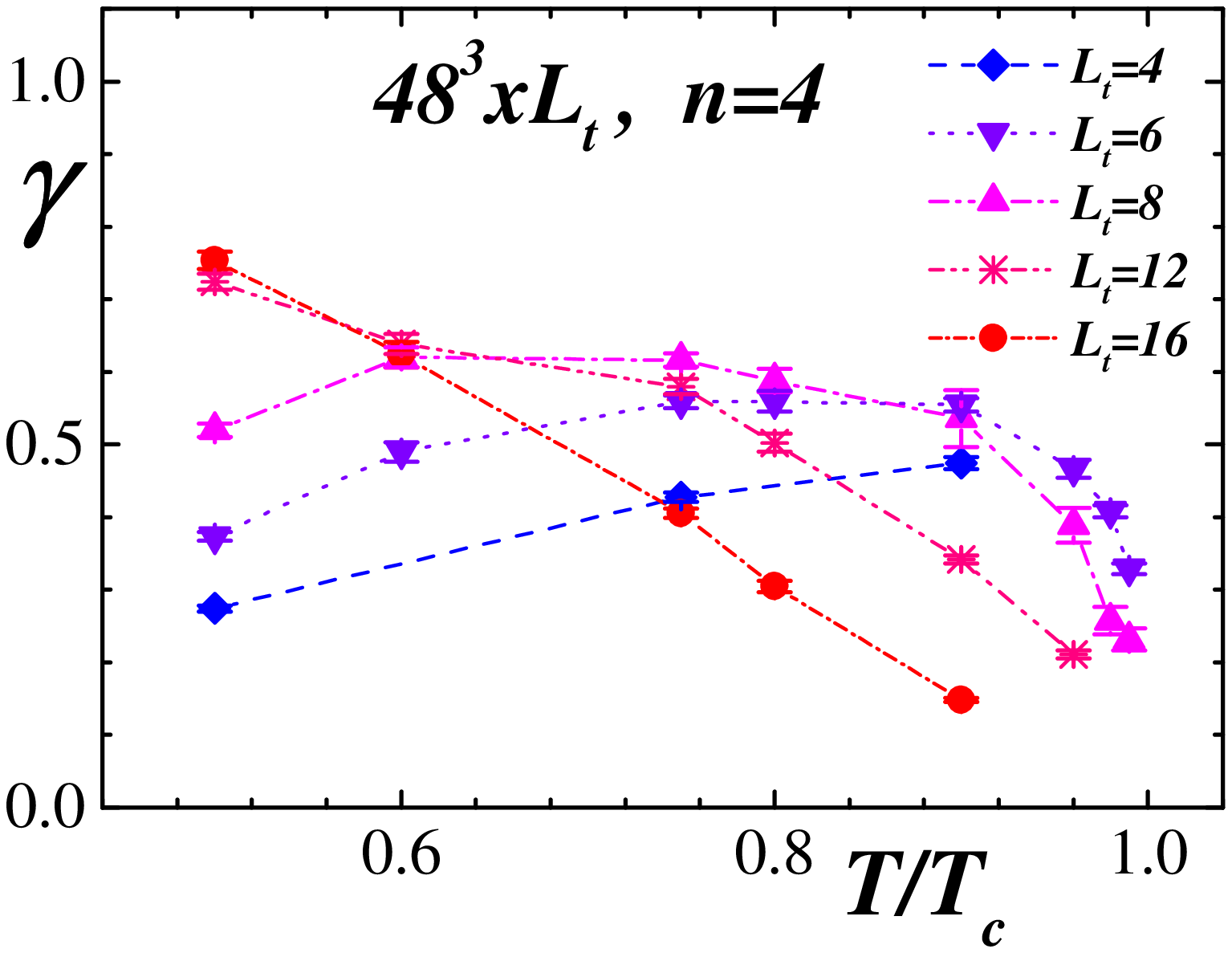} \vspace{-2mm}\\
(c) & (d)
\end{tabular}
\end{center}
\vspace{-3mm}
\caption{The coefficient $\gamma$ {\it vs.} temperature $T$ for various temporal extensions of the lattice
and for various blacking factors $n$.}
\label{fig:gamma:T}
\end{figure}
\begin{figure}[!htb]
\begin{center}
\vspace{3mm}
\begin{tabular}{cc}
\includegraphics[scale=0.45,clip=true]{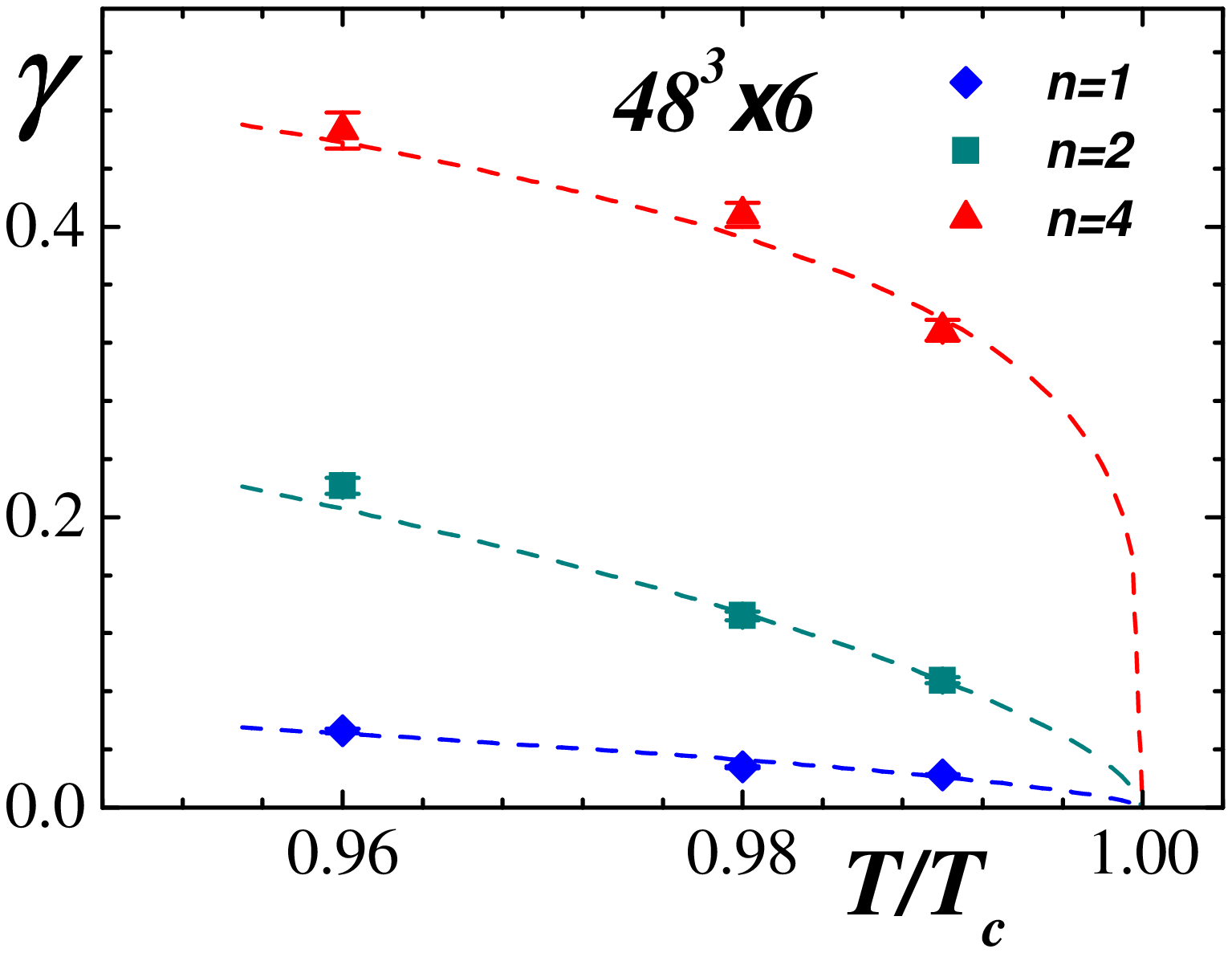} \hspace{5mm} &
\includegraphics[scale=0.45,clip=true]{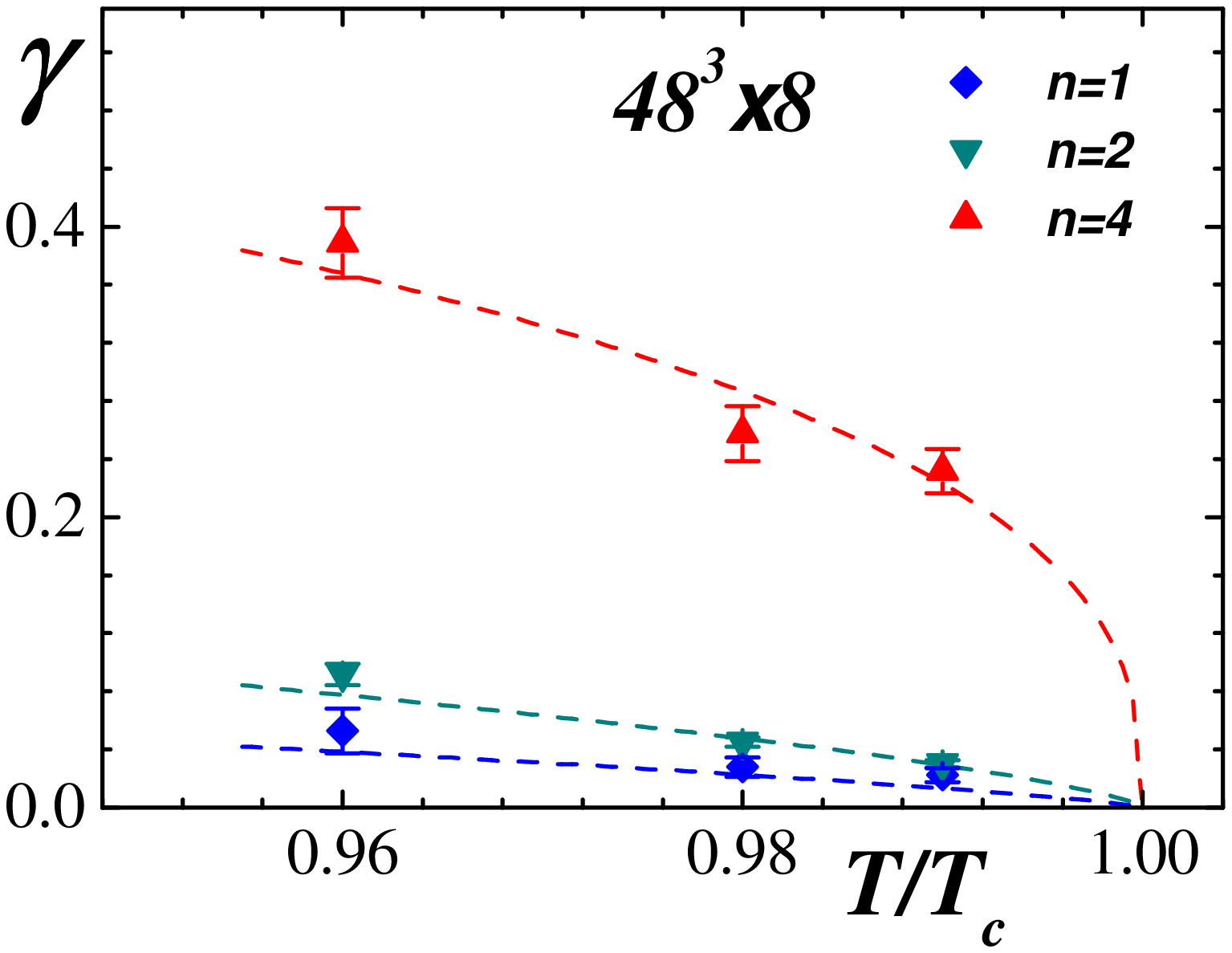} \\
(a) & (b)
\end{tabular}
\end{center}
\vspace{-3mm}
\caption{Examples of the fits of the $\gamma$ parameter as the function of the temperature.}
\label{fig:gamma:fits}
\end{figure}

At the critical temperature, $T=T_c$, the $4D$ IR monopole cluster disappears and we expect
a similar behavior for the $3D$ projected IR cluster. This implies, that the parameter $\gamma(b,T)$
must become a (non-local) {\it order parameter}: it must vanish at the critical point.
One can reach this conclusion by noticing that the parameter $\gamma(b,T)$ is proportional to
the monopole density~\eq{eq:density} (which vanishes at $T=T_c$), and that the factor $A$ is
unlikely to be divergent at the critical temperature (what can also be deduced from Figures~\ref{fig:A}).

We show that the quantity $\gamma$ is indeed an order parameter in
Figure~\ref{fig:gamma:T}(a) which depicts $\gamma$ for elementary ($n=1$)
monopoles as a function of temperature for various temporal extensions of the lattice. The behavior of the
$\gamma$-parameter in the vicinity of the phase transition point depends on the value of the temporal extension
$L_t$. However, the value of the $\gamma$--parameter at the critical temperature, $\gamma(T \to T_c) \to 0$,
is universal with respect to $L_t$. Moreover, one can observe in
Figures~\ref{fig:gamma:T}(b,c,d)  -- which correspond to the extensions $n=2,3,4$, respectively -- that the
$\gamma$-coefficient for the extended monopoles is also vanishing at $T=T_c$.

To characterize the critical behavior of the parameter $\gamma$ in the vicinity of the phase transition
we have fitted this parameter by the function
\beqn
\gamma^{\mathrm{fit}}(b,T) = C_\gamma \cdot \Bigl( 1 - \frac{T}{T_c}\Bigr)^\delta\,, \qquad T<T_c\,,
\label{eq:delta:fit}
\eeqn
where $\delta$ and $C_\gamma$ are the fitting parameters. We performed the fits for various lattices $L_s^3\times L_t$
and extensions $n$. The results for the "critical exponent" $\delta$ are shown in Table~\ref{tbl:delta}.
\begin{table}
\begin{center}
\begin{tabular}{|c|c|c|c|}
\hline
\multicolumn{4}{|c|}{$\delta$} \\
\hline
$\ n\ $& $\ 48^3\times6\ $ & $\ 48^3\times8\ $ & $\ 72^3\times8\ $ \\
\hline
1  & 0.64(15) & 0.76(2)   &  --      \\
2  & 0.62(8)  & 0.70(16)  & 0.48(3)  \\
3  & 0.34(6)  & 0.55(7)   & 0.30(2)  \\
4  & 0.22(2)  & 0.36(6)   & 0.18(3)  \\
6  & 0.11(2)  & 0.20(2)   &  --      \\
\hline
\end{tabular}
\end{center}
\caption{The "critical exponents" $\delta$ -- obtained with the help of the fit~\eq{eq:delta:fit} --
for various lattices $L_s^3\times L_t$ and extensions $n$.}
\label{tbl:delta}
\end{table}
{}From this table one notices that the quantity $\delta$ is not universal: it depends not only on the extension of the monopole
blocking but also it also depends on the lattice volume.
Moreover, the larger extension $n$ the steeper behavior of $\gamma$ in the vicinity
of the phase transition is.

One should add a word of caution here. The fit results shown in Table~\ref{tbl:delta} are crucially dependent of the $T/T_c = 0.98,0.99$
points (as one can see from Figures~\ref{fig:gamma:fits}(a),(b)), which are very close to the phase transition.
Since the transition is of the second order then the finite--volume effects must be strong and the results of fits may quantitatively be
incorrect (although the results presented in Figures~\ref{fig:gamma:fits}(a),(b) must qualitatively be correct).

\section{Monopole entropy}
\label{sec:entropy}

Apart from the finite--volume effect, the distribution~\eq{eq:IR:distr:two}
has contributions from the energy and the entropy. As  seen above,
the action contribution is proportional to $e^{- f_0 L}$. The entropy
contribution is proportional to $\mu^L$ (with $\mu>0$) for sufficiently
large monopole lengths, $L$. Thus, the entropy factor, $\mu$, is
\beqn
\mu = \exp\{f_0 + \gamma\}\,.
\label{eq:mu}
\eeqn

\begin{figure}[!htb]
\begin{center}
\begin{tabular}{cc}
\includegraphics[scale=0.47,clip=true]{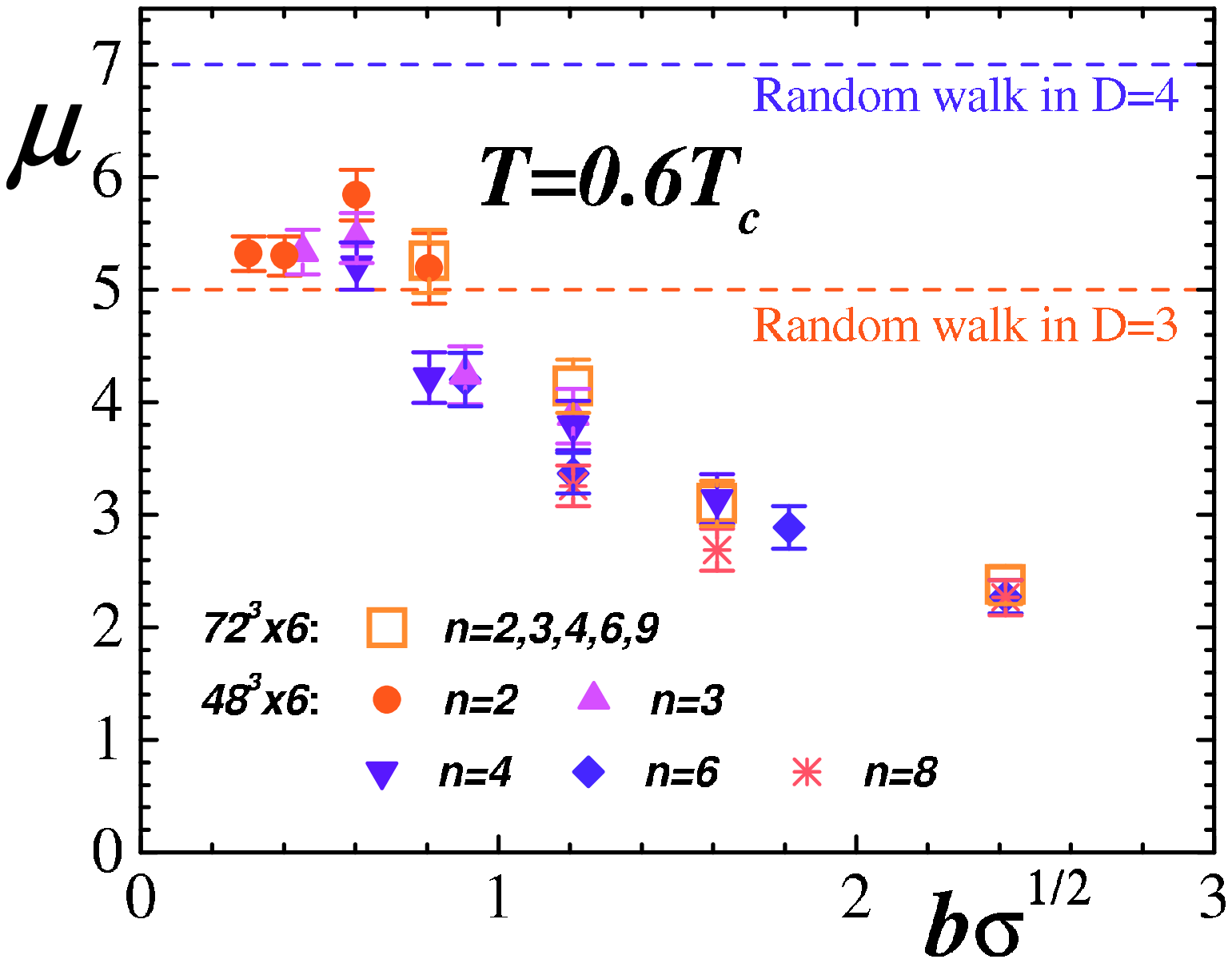} \hspace{5mm} &
\includegraphics[scale=0.47,clip=true]{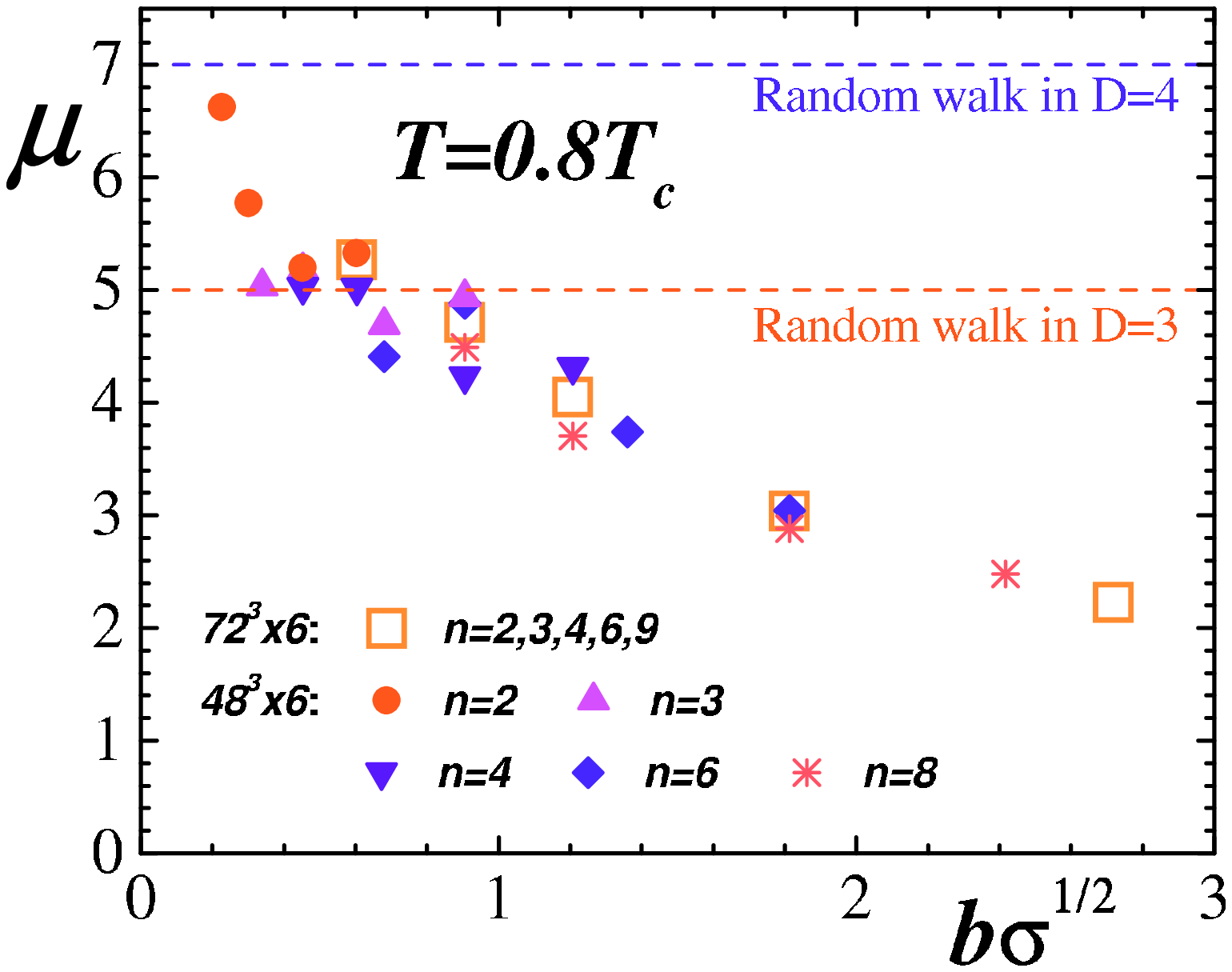} \\
(a) & (b) \vspace{7mm} \\[2mm]
\includegraphics[scale=0.47,clip=true]{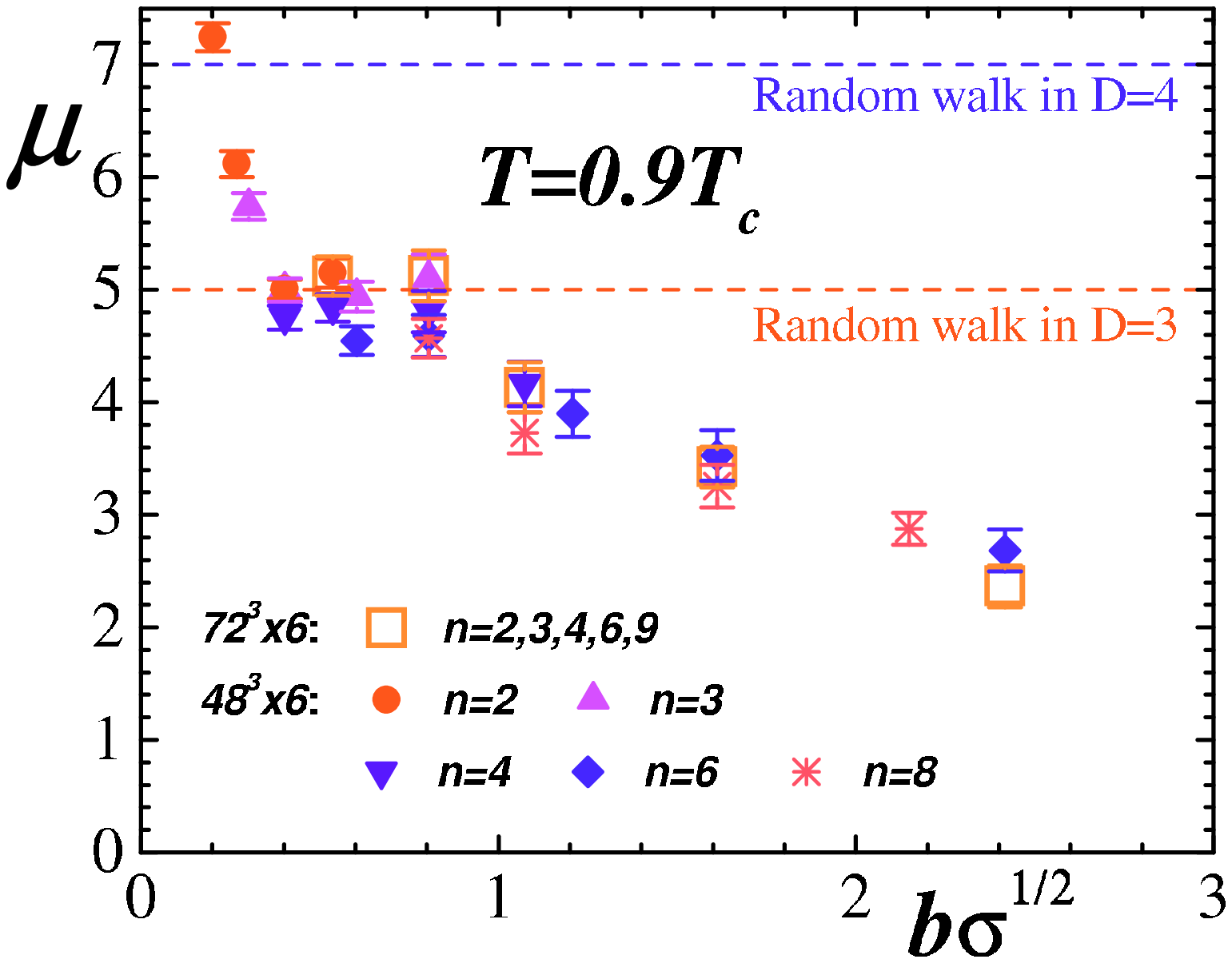} \hspace{5mm} &
\includegraphics[scale=0.47,clip=true]{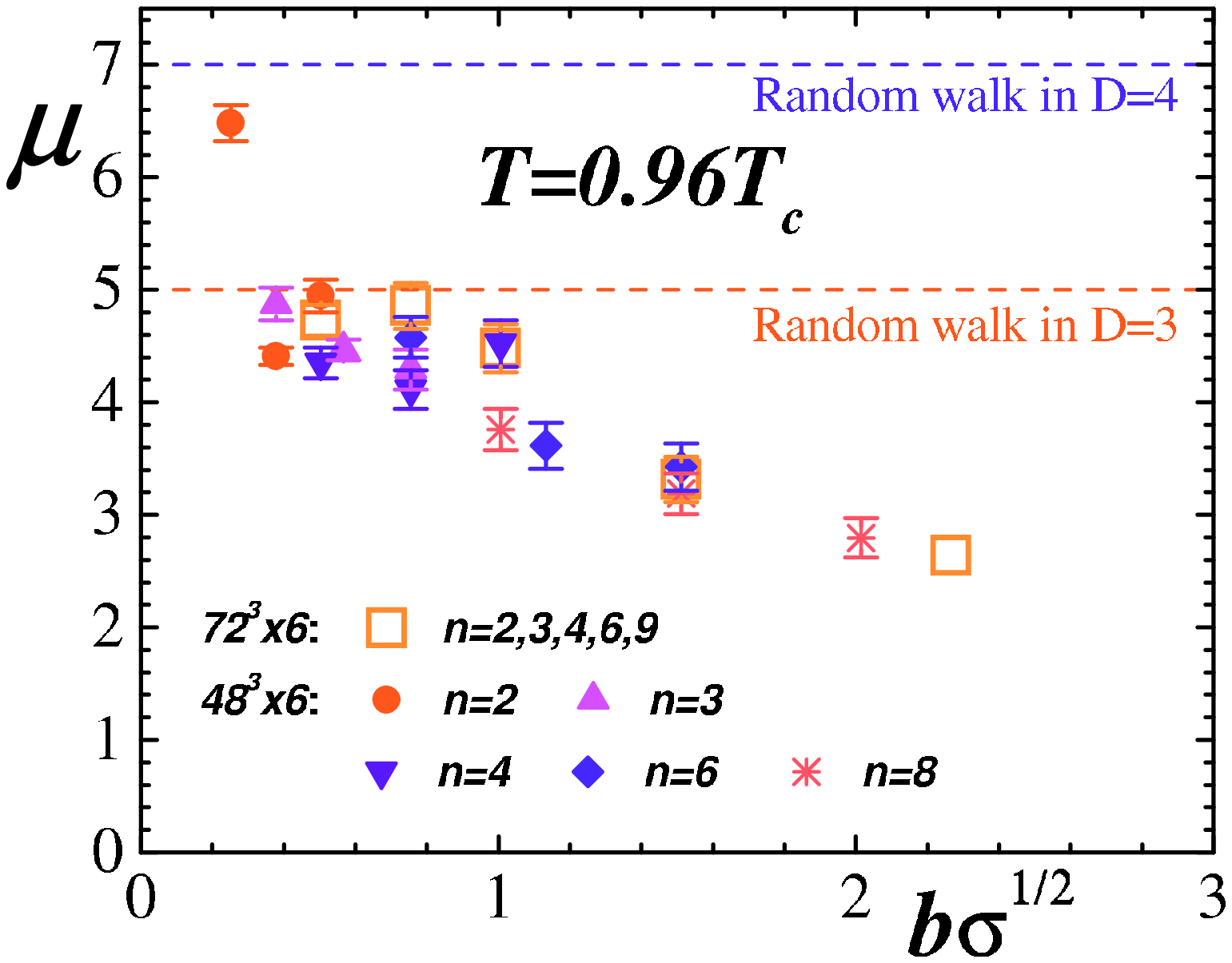} \\
(c) & (d) \vspace{7mm} \\[2mm]
\includegraphics[scale=0.47,clip=true]{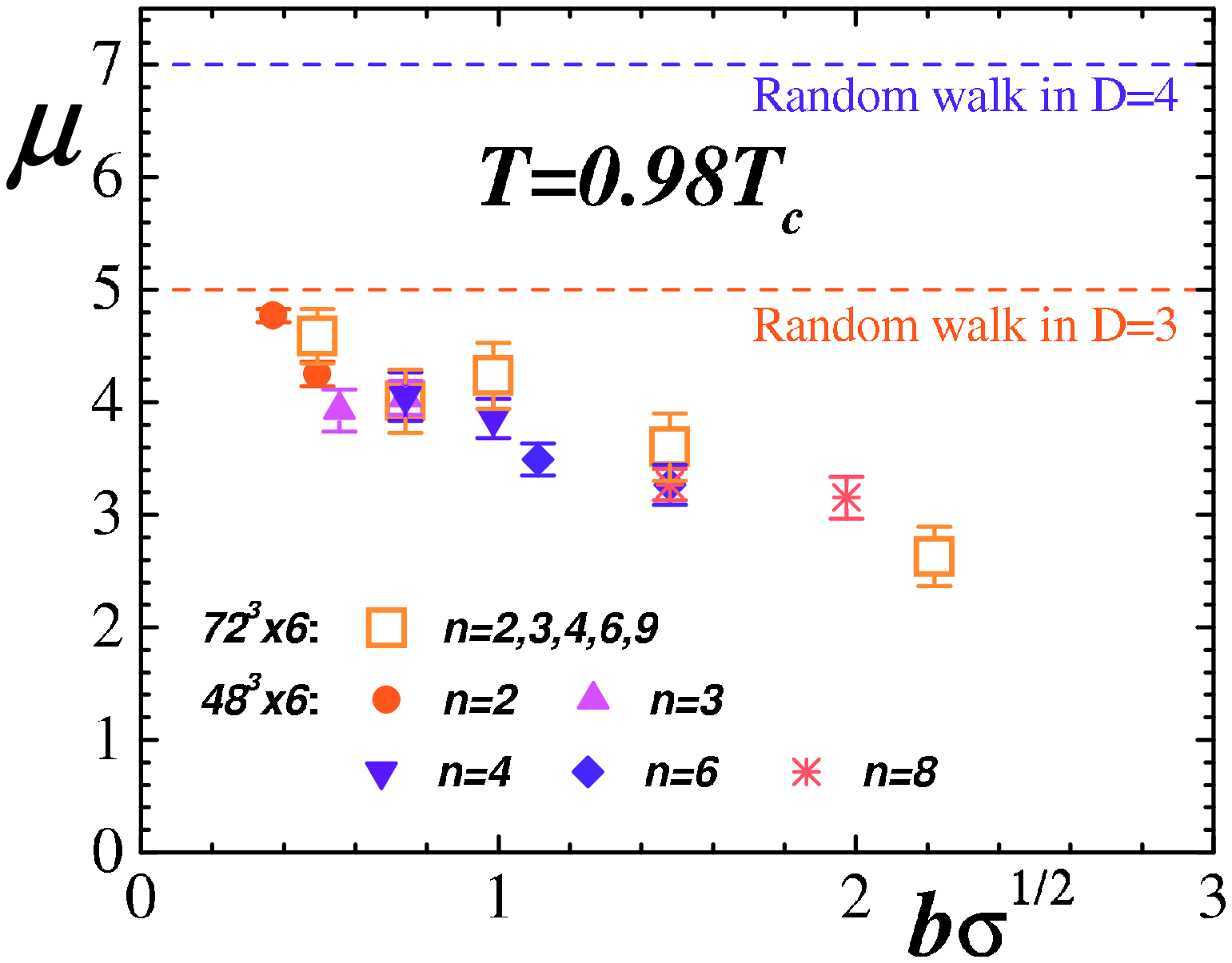} \hspace{5mm} &
\includegraphics[scale=0.47,clip=true]{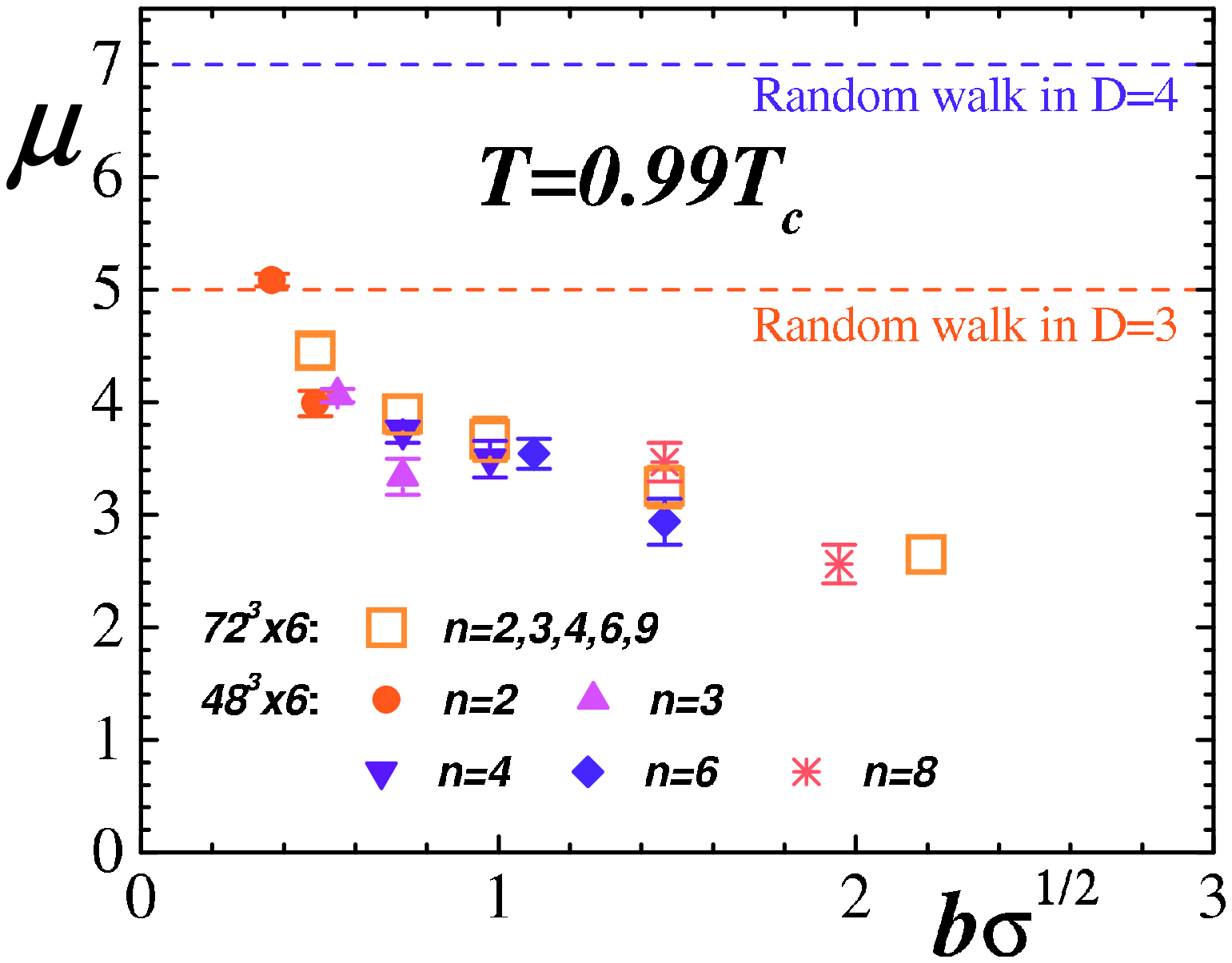} \\
(e) & (f)
\end{tabular}
\end{center}
\caption{Entropy factor of the spatially projected monopole currents as the function
of the scale $b$ at various temperatures.}
\label{fig:entropy}
\end{figure}

We determine the entropy using Eq.~\eq{eq:mu}. The numerical results
for the entropy factor $\mu(b,T)$ are shown in Figures~\ref{fig:entropy} for
various temperatures, lattice volumes and blocking factors.
One can see that the entropy factor $\mu$ scales as the function of $b$, as expected.

In order to understand the meaning of the data shown in Figure~\ref{fig:entropy} we note that
if the monopoles are randomly walking on a $3D$
hypercubic lattice then we should get a definite value for the entropy factor, $\mu=5$. This is because
at each site there exist five choices for the monopole current to go further. One can see that
far from the phase transition, $T \lesssim 0.96 T_c$
the entropy factor $\mu$ indeed tends to the $\mu=5$ plateau at moderately small values of $b \sim 0.4 \dots 1$.
At yet smaller $b$ the entropy gets bigger than random walk value $\mu>5$
because in this region the inverse Monte-Carlo method with the truncated quadratic monopole action
does not work well~\cite{chernodub}. Thus, the value of the constant $f_0$ -- defined in Eq.~\eq{eq:length:proportionality} --
can not be obtained correctly.

At large $b$ the entropy factor drops down with the increase of the factor $b$. This feature is
independent of the temperature. In the zero temperature case~\cite{IshiguroSuzuki}
the entropy factor $\mu$ approaches unity in the $b \to \infty$ limit. This feature is difficult to
observe from our data since the information about the entropy at large values of the blocking size $b$ is not available.

\section{Conclusion}

The distributions of the spatially--projected infrared monopole currents of various blocking sizes,
$n$ were studied on the lattices with different spacings, $a$, and volumes, $L_s^3\times L_t$.
We find that the distributions can be described by a gaussian anzatz with a good accuracy. The anzatz contains two
important terms: (i) the linear term, which contains information about the energy and entropy of the monopole
currents; and (ii) the quadratic term, which appears due to finite--volume and which suppresses large
infrared clusters. The linear term is independent of the lattice volume while the quadratic term is
inversely proportional to the volume. Moreover, the linear term is a (non--local) order parameter for the
deconfinement phase transition.

To get the entropy of the spatially--projected currents we studied the action of the monopoles
belonging to the infrared monopole clusters of the  spatially--projected currents
using an inverse Monte-Carlo method. We show that the entropy factor has a plateau
at sufficiently small values of $b$ and at $T \lesssim 0.96 T_c$. A reason for the temperature
restriction of our result is that our analysis may not be valid close to the second order phase transition point
because of the increase of correlation lengthes at (and, consequently, because of
strong finite-volume effects) $T \approx T_c$.  At $b \gtrsim 1$ the entropy is a descending
function of $b=n a$, indicating that the effective degrees of freedom of the projected and blocked monopoles are getting
smaller as the blocking scale $b$ increases. This effect is very similar to the zero temperature case, in which the
monopole motion corresponds to the classical picture: the monopole
with the large blocking size $b$ becomes a macroscopic object and the motion of such a monopole gets
close to a straight line.

\clearpage

\begin{acknowledgments}
M.N.Ch. is supported by grants RFBR 04-02-16079, MK-4019.2004.2
and by JSPS Grant-in-Aid for Scientific Research (B) No.15340073.
T.S. is partially supported by JSPS Grant-in-Aid for
Scientific Research on Priority Areas No.13135210 and (B)
No.15340073. This work is also supported by the Supercomputer
Project of the Institute of Physical and Chemical Research
(RIKEN). A part of our numerical simulations have been done using
NEC SX-5 at Research Center for Nuclear Physics (RCNP) of Osaka
University. M.N.Ch. is grateful to the members of Institute for
Theoretical Physics of Kanazawa University for the kind
hospitality and stimulating environment.
\end{acknowledgments}

\end{document}